\documentstyle[aps,pra,multicol,epsfig]{revtex}
\def\vecgr#1{\mathchoice{\mbox{\boldmath$\mathrm\displaystyle#1$}}
{\mbox{\boldmath$\mathrm\textstyle#1$}}
{\mbox{\boldmath$\mathrm\scriptstyle#1$}}
{\mbox{\boldmath$\mathrm\scriptscriptstyle#1$}}}
\def\vec#1{\mathchoice{\mathrm{\mathbf{\displaystyle#1}}}
{\mathrm{\mathbf{\textstyle#1}}}
{\mathrm{\mathbf{\scriptstyle#1}}}
{\mathrm{\mathbf{\scriptscriptstyle#1}}}}

\newsavebox{\LSIM}
\sbox{\LSIM}{\raisebox{-1ex}{$\ \stackrel{\textstyle<}{\sim}\ $}}
\newcommand{\lsim}{\usebox{\LSIM}}
\newsavebox{\GSIM}
\sbox{\GSIM}{\raisebox{-1ex}{$\ \stackrel{\textstyle>}{\sim}\ $}}
\newcommand{\gsim}{\usebox{\GSIM}}

\newcommand{\be}{\begin{equation}}
\newcommand{\ee}{\end{equation}}
\newcommand{\bea}{\begin{eqnarray}}
\newcommand{\eea}{\end{eqnarray}}

\newcommand{\lk}{\left}
\newcommand{\rk}{\right}
\newcommand{\ra}{\,\rangle}
\newcommand{\la}{\langle\,}
\newcommand{\psidg}{\psi^\dagger}

\newcommand{\Tr}{\mbox{Tr$\,$}}

\newcommand{\eh}{\mbox{$\frac{1}{2}$}}

\newcommand{\del}{\delta}

\newcommand{\cN}{{\cal N}}
\newcommand{\cO}{{\cal O}}
\newcommand{\cT}{{\cal T}}

\newcommand{\nabv}{\vecgr{\nabla}}
\newcommand{\pv}{\vec{p}}
\newcommand{\rv}{\vec{r}}
\newcommand{\Rv}{\vec{R}}
\newcommand{\xv}{\vec{x}}
\newcommand{\yv}{\vec{y}}

\newcommand{\ptdnd}[1]{\frac{\partial}{\partial #1}}
\newcommand{\delndel}[1]{\frac{\delta}{\delta #1}}
\newcommand{\intdd}[1]{\int\,d^3\!#1\,}

\newcommand{\intddz}[2]{\int\,d^3\!#1\,d^3\!#2\,}
\newcommand{\intddd}[3]{\int\,d^3\!#1\,d^3\!#2\,d^3\!#3\,}
\newcommand{\intddv}[4]{\int\,d^3\!#1\,d^3\!#2\,d^3\!#3\,d^3\!#4\,}

\newcommand{\kfrac}[2]{\mbox{$\frac{#1}{#2}$}}

\newcommand{\oneB}{{\mathrm 1B}}
\newcommand{\twoB}{{\mathrm 2B}}

\begin{document}
\draft
\title{Microscopic theory of atom-molecule oscillations in a
Bose-Einstein condensate}
\author{Thorsten K\"ohler, Thomas Gasenzer,
and Keith Burnett}
\address{Clarendon Laboratory, Department of Physics, University of Oxford,
Oxford OX1~3PU, United Kingdom\\(\today)}
\maketitle

\begin{abstract}
\noindent
In a recent experiment at JILA 
[E.A.~Donley et al., Nature (London) {\bf 417}, 529 (2002)]
an initially pure condensate of $^{85}$Rb atoms was exposed to a specially
designed time dependent magnetic field pulse in the vicinity of a Feshbach 
resonance. The production of new components of the gas as well as their 
oscillatory behavior have been reported. We apply a microscopic 
theory of the gas to identify these components and determine 
their physical properties. Our time dependent studies 
allow us to explain the observed dynamic evolution of all fractions, and to 
identify the physical relevance of the pulse shape.
Based on ab initio predictions, our theory strongly supports the view
that the experiments have produced a molecular condensate.

\end{abstract}
\begin{multicols}{2}

%
%
\setcounter{equation}{0}
\section{Introduction}
\label{sec1}
\noindent
The subject of the coupling between atoms and molecules in Bose-Einstein
condensates has attracted intense interest following recent experiments
\cite{Stenger99,Wynar00,Claussen02,Donley02}.
Particular experiments we shall focus on are those performed at JILA 
\cite{Claussen02,Donley02} 
where the strength of the inter-atomic potential of $^{85}$Rb was varied 
rapidly using specially designed magnetic field pulses. This resulted in the 
loss of condensate atoms and the production of new
components in the gas. One of these components is believed to be composed
of molecules, and to be a molecular condensate. This is a remarkable
achievement, with profound consequences for future work in the field. We
shall show that this interpretation is fully supported by the theoretical
work described in this article. We want to emphasize that the prediction of a
molecular condensate arises naturally from the theory and does not have to
be assumed at the outset. To make this prediction we use the microscopic
theory of evolving condensed systems developed in 
Ref.~\cite{Koehler02}. This theory allows
us to include the full dynamics of colliding pairs of atoms without the need
for any assumptions about the nature of the states produced in the
experiment. This theory gives us a generalization of the well known
Gross-Pitaevskii equation (GPE) which includes the binary dynamics fully 
in the description of time evolving condensates.

If the variation in the magnetic field occurs slowly in comparison with the
duration of a collision one should expect to be able to use the standard
Gross-Pitaevskii approach 
\cite{Dalfovo99}
to the problem. The derivation of the GPE, however, relies precisely on the 
assumption that collisions occur on a
timescale small compared to all others in the problem
\cite{Koehler02}. This approximation,
therefore, fails in this new experimental regime where the magnetic field,
tuned in the region of a Feshbach resonance, varies on this timescale. 

As mentioned above the interpretation of the results of the experiment posits
the production of bound molecular states, persisting at the end of the
magnetic pulse sequence. Some theoretical treatments of the problem of
molecules in condensates 
\cite{Kokkelmans02,Mackie02}
separate out such states as a separate entity of the physical system at the 
outset of their calculation, i.e.~physical observables associated 
with two-body bound states are described in terms of a molecular quantum 
field that emerges directly in a model Hamiltonian
\cite{Timmermans99}.  

In this article we give a microscopic
treatment of the evolution of a Bose-Einstein condensate in the presence of
a time varying magnetic field that completely avoids assumptions on the
nature of the states involved in the collision process. In fact we treat the
binary events involving bound and free molecular states formed during the
evolution in a unified manner. This is a most sensible approach as strictly
speaking, in the presence of the non-adiabatically time varying field there 
is no proper distinction between bound and free states. 
At the end of the pulse sequence
we can of course resolve the final state of the gas into free and bound
components. To do this we only use the assumption that the gas remains
dilute and that binary encounters are the dominant collisional process. The
evolution of pairs of particles from the condensate into other free states
or into bound molecular states comes from this treatment. Our theory
strongly supports the view that the experiments have produced a molecular
condensate. We should emphasize again that this conclusion comes from an ab
initio prediction of the theory and not as an assumption. 

In the following sections we review briefly the microscopic theory of a
dilute gas that we use in the analysis of the problem. We then show how the
macroscopic evolution of the condensates is coupled in and out of the binary
dynamics. We can then produce explicit expressions for the various
components that are produced in the experiments. We have performed
calculations, both for the case of the homogeneous gas and also for the case
of a trapped condensate. The qualitative results of these two calculations
agree but there are quantitative differences that merit further study. Our
results for the loss of condensate and the production of a heated component
agree with those produced in the experiment. In addition we are able to
confirm the presence of a molecular condensate with all the physical
properties we would expect. 


%
%
\section{Microscopic dynamics approach}
\label{sec2}
%
%
\subsection{Atomic mean field}
\label{sec2.2}
\noindent
The microscopic dynamics approach 
\cite{Koehler02}
is based on the general many body Hamiltonian for identical bosons with
a pair interaction $V(\rv,t)$,
\bea
 &&H
 =\intdd{x}\psidg(\xv)H_{\mathrm 1B}(\xv)\psi(\xv)
 \nonumber\\
 &&+\ \frac{1}{2}\intddz{x}{y}\,
     \psidg(\xv)\psidg(\yv)V(\xv-\yv,t)\psi(\yv)\psi(\xv).
\label{eq2.2}
\eea
Here, $H_{\mathrm 1B}(\xv)=-\hbar^2\nabv^2/2m+V_{\mathrm trap}$ is the  
Hamiltonian of a single atom containing the kinetic energy 
and the trapping potential.
The field operators satisfy bosonic commutation
relations, $[\psi(\xv_1),\psi(\xv_2)]=0$, and
$[\psi(\xv_1),\psidg(\xv_2)]=\del(\xv_1-\xv_2)$.
In the situation studied in this article, the inter-atomic interaction
is varied using an external magnetic field pulse in the vicinity of a Feshbach
resonance 
\cite{Donley02}
and its time dependence is noted explicitly in Eq.~(\ref{eq2.2}).

All physical properties of a gas of atoms can be determined from  
correlation functions, $\la\psidg(\xv_n)\cdots\psi(\xv_1)\ra_t$,
i.e.~expectation values of normal ordered products
of field operators with respect to the quantum state of the gas at time $t$.
References \cite{Fricke96,Koehler02} provide a general scheme to transform the 
exact infinite hierarchy of coupled dynamic equations for correlation 
functions into a more favorable form: The resulting equivalent set of dynamic 
equations for what are called non-commutative cumulants allows for a 
systematic truncation in accordance with Wick's theorem in statistical 
mechanics. In this article we apply this truncation scheme to determine a 
closed set of equations of motion for the relevant physical quantities. The 
derivation of the approach to the level of approximation required to study the 
phenomena reported in 
\cite{Donley02}, i.e.~the first order microscopic dynamics approach
\cite{Koehler02}, is given in Appendix \ref{appA}.

The relevant physical quantities involve only the first and second order 
cumulants:
\bea
\label{eq2.7}
  \Psi(\xv,t)
  &=& \la\psi(\xv)\ra_t,
  \nonumber\\
  \Phi(\xv,\yv,t)
  &=& \la\psi(\yv)\psi(\xv)\ra_t-\la\psi(\yv)\ra_t\la\psi(\xv)\ra_t,
  \nonumber\\
  \Gamma(\xv,\yv,t)
  &=&
  \la\psidg(\yv)\psi(\xv)\ra_t-\la\psidg(\yv)\ra_t\la\psi(\xv)\ra_t.
\eea
Here, $\Psi(\xv,t)$ is the atomic mean field, $\Phi(\xv,\yv,t)$
the pair function, which plays an important role in the description of 
correlated pairs of atoms, and $\Gamma(\xv,\yv,t)$ is the one-body density
matrix of the non-condensed fraction.
The density of the gas at the position $\xv$ and time $t$ is thus given by
$n(\xv,t)=\la\psidg(\xv)\psi(\xv)\ra_t=\Gamma(\xv,\xv,t)+|\Psi(\xv,t)|^2$.

In the first order microscopic dynamics approach the atomic mean field is
determined through a closed nonlinear Schr\"odinger equation
\cite{Koehler02}:
\bea
\label{eq2.20}
  &&i\hbar\ptdnd{t}\Psi(\xv,t)
  = H_{\mathrm 1B}(\xv)\Psi(\xv,t)
  \nonumber\\
  &&\quad-\ \Psi^*(\xv,t)\int_{t_0}^\infty d\tau
  \Psi^2(\xv,\tau)\frac{\partial}{\partial \tau}h(t,\tau).
\eea
The collision term distinguishes the non-Markovian dynamic 
Eq.~(\ref{eq2.20}) from the Gross-Pitaevskii approach and is determined
through the coupling function 
\be
\label{eq3.9}
  h(t,\tau)
  =(2\pi\hbar)^3\la0\,|\,V(t)U_\twoB(t,\tau)\,|\,0\ra\theta(t-\tau),
\ee
where $U_\twoB(t,\tau)$ denotes the unitary time development operator of the
relative motion of two atoms in free space, 
$|0\ra$ is the zero momentum plane wave and $\theta(t-\tau)$ 
is the step function which gives unity for $t>\tau$ and vanishes elsewhere. 
Throughout this article the three dimensional plane wave with momentum $\pv$ 
is normalized as 
$\la \rv|\pv\ra = \exp(i\pv\cdot\rv/\hbar)/\sqrt{2\pi\hbar}^3$.
%
%
\subsection{Non-condensed fraction}
\label{sec2.4}
\noindent
In the first order microscopic dynamics approach the nonlinear Schr\"odinger
Eq.~(\ref{eq2.20}) determines not only the atomic mean field but also the 
pair function, $\Phi$, and, in turn, the density matrix of the non-condensed 
fraction, $\Gamma$, in Eq.~(\ref{eq2.7}). The pair function is given by
\bea
   \nonumber
   \Phi(\xv,\yv,t)
   =&&-\int_{t_0}^td\tau \intddz{x'}{y'}\Psi(\xv',\tau)\Psi(\yv',\tau)\\
   &&\times\frac{\partial}{\partial \tau}
   \la\xv,\yv|U_{\rm trap}^{\rm 2B}(t,\tau)|\xv',\yv'\ra,
   \label{eq4.11}
\eea
where $U_{\rm trap}^{\rm 2B}(t,\tau)$ is the unitary time development 
operator of two trapped atoms interacting through the pair potential
$V(t)$. 
The density matrix of the non-condensed fraction expressed in terms of the
pair function is given by
\be
\label{eq4.112}
\Gamma({\bf x},{\bf y},t)=\intdd{x}'\Phi({\bf x},{\bf x}',t)
\Phi^*({\bf y},{\bf x}',t). 
\ee
As shown in Appendix \ref{appA}, Eq.~(\ref{eq4.112}) assures both the 
positivity of all occupation numbers and the conservation of the total 
number of atoms in the gas:
\be
\label{eq4.113}
\intdd{x}\left[|\Psi({\bf x},t)|^2+\Gamma({\bf x},{\bf x},t)\right]=
N_{\rm c}(t)+N_{\rm nc}(t)=N.
\ee

The form of Eq.~(\ref{eq4.112}) suggests a separation of the number
of non-condensed atoms into a molecular fraction and correlated pairs
of atoms after a time dependent magnetic field pulse of the kind reported
in Ref.~\cite{Donley02} as follows: The total number of non-condensed 
atoms is given by
\bea
\nonumber
N_{\rm nc}(t)&=&\intdd{x}\Gamma({\bf x},{\bf x},t)=
\intddz{x}{x'}|\Phi({\bf x},{\bf x}',t)|^2\\
&=&\intddz{R}{r}|\Phi({\bf R},{\bf r},t)|^2,
\label{eq4.114}
\eea
where the position dependence was changed to two body center of mass and
relative coordinates ${\bf R}=({\bf x}+{\bf y})/2$ and 
${\bf r}={\bf y}-{\bf x}$, respectively.
Under the assumption that the trap is switched off at time $t_{\rm fin}$,
immediately after the pulse, and the magnetic field is held constant at its
final value, the energy states of the relative motion of a pair of atoms 
become stationary. A complete set of energy eigenstates is given through
\be
1=\sum_\nu|\phi_{{\rm b}\nu}\ra\la\phi_{{\rm b}\nu}|
+\intdd{p}|\phi_{\pv}^{(+)}\ra\la\phi_{\pv}^{(+)}|,
\label{eq4.115}
\ee  
where $\phi_{{\rm b}\nu}$ are the molecular bound states of the final pair 
potential and $\phi_{\pv}^{(+)}$ are chosen as stationary scattering 
states which, at large relative distance, become a sum of an incoming plane
wave with momentum $\pv$ and an outgoing spherical wave 
(see, e.g., \cite{Messiah} or Appendix \ref{appB}). 
Replacing the spatial integration over the relative 
coordinate $\rv$ in Eq.~(\ref{eq4.114}) in favor of the energy eigenstates 
in Eq.~(\ref{eq4.115}) the non-condensed fraction splits into a molecular part
and a scattering part:
\bea
   \nonumber
   N_{\rm nc}(t)
   =\intdd{R}\bigg[&&\sum_\nu\,|\la\Rv,\phi_{{\rm b}\nu}\,|\,\Phi(t)\ra|^2\\
    && +\intdd{p}|\la\Rv,\phi_{\pv}^{(+)}\,|\Phi(t)\ra|^2\bigg],
\label{eq4.12}
\eea
where $|\,\Phi(t)\ra\equiv\intddz{R}{r}|\,\Rv,\rv\ra\Phi(\Rv,\rv,t)$. 
The choice of eigenstates and the physical meaning of the contributions to
Eq.~(\ref{eq4.12}) depend on the experimental situation to be described. 
As will be shown in the next subsections the molecular part in 
Eq.~(\ref{eq4.12})
determines the number of atoms bound to molecules after the pulse while the 
scattering part describes pairs of atoms emitted from the condensate in a 
ballistic expansion.
%
%
\subsection{Molecular fraction}
\label{sec2.3}
\noindent
The operator that determines the number of pairs of atoms in the 
specific bound state $\phi_{\rm b}$ in a gas with $N$ atoms reads,  
in its first quantization form,
\be
\label{eq4.0}
  \cN_{\rm b} = \frac{1}{2}\sum_{{i,j=1}\atop{i\not=j}}^N
\,|\,\phi_{{\rm b},ij}\ra\la\phi_{{\rm b},ij}\,|,
\ee
where $i$ and $j$ indicate the pair of atoms. Expressed in terms of the 
atomic field operator Eq.~(\ref{eq4.0}) becomes:
\bea
\label{eq4.1}
  \cN_{\rm b}
  &=& \frac{1}{2}\intddv{x_1}{x_2}{x_1'}{x_2'}
    \phi_{\rm b}(\xv_2'-\xv_1')
    \phi_{\rm b}^*(\xv_2-\xv_1)
  \nonumber\\
  &&\quad\times\ \delta\lk(\eh(\xv_1'+\xv_2')-\eh(\xv_1+\xv_2)\rk)
  \nonumber\\
  &&\quad\times\ \psidg(\xv_1')\psidg(\xv_2')\psi(\xv_2)\psi(\xv_1).
\eea
The mean number of molecules in the state $|\,\phi_{\rm b}\ra$ is thus 
given by
\bea
\label{eq4.2}
   &&N_{\rm b}(t)\ =\ \la\cN_{\rm b}\ra_t
   \ = \ \frac{1}{2}\intddd{r'}{r}{R}\phi_{\rm b}(\rv')\phi_{\rm b}^*(\rv)
   \nonumber\\
   &&\ \times\
   \lk\la
   \psidg(\Rv+\kfrac{\rv'}{2})\psidg(\Rv-\kfrac{\rv'}{2})
   \psi(\Rv-\kfrac{\rv}{2})\psi(\Rv+\kfrac{\rv}{2})\rk\rangle_t,
\eea
where $\Rv$ and $\rv$ are center of mass and relative coordinates, 
respectively.
The fourth order correlation function in Eq.~(\ref{eq4.2}) can be
factorized into cumulants (cf.~Eq.~(\ref{eq2.6})), and truncated
in accordance with the level of approximation of the first order microscopic 
dynamics approach:
\bea
\label{eq4.2a}
  &&\lk\la\psidg(\xv_4)\psidg(\xv_3)\psi(\xv_2)\psi(\xv_1)\rk\rangle_t
  \nonumber\\
  &&=\ \lk\la\psidg(\xv_4)\psidg(\xv_3)\rk\rangle_t
      \big\la\psi(\xv_2)  \psi(\xv_1) \big\rangle_t.
\eea
The mean number of molecules in the state $|\,\phi_{\rm b}\ra$
can then be expressed in terms of a molecular mean field as
\be
\label{eq4.3}
  N_{\rm b}(t)
  = \intdd{R}|\Psi_{\rm b}(\Rv,t)|^2,
\ee
where
\bea
  &&\Psi_{\rm b}(\Rv,t)
  \ =\ \frac{1}{\sqrt{2}}\intdd{r}\phi_{\rm b}^*(\rv)
  \nonumber\\
  &&\quad\times\ \lk[
     \Phi(\Rv,\rv,t)
    +\Psi(\Rv+\kfrac{\rv}{2},t)\Psi(\Rv-\kfrac{\rv}{2},t)\rk].
\label{eq4.4}
\eea
The overlap of the molecular wave function $\phi_{\rm b}$ with the second, 
factorized term on the right hand side of Eq.~(\ref{eq4.4}) can be shown 
to be negligible in all applications described in this article. 
The molecular part on
the right hand side of Eq.~(\ref{eq4.12}) is thus twice the number of 
dimer molecules in the gas, i.e.~the number of atoms bound to dimer molecules.
The wave function 
$\Psi_{\rm b}(\Rv,t)$, which yields the density of the molecular fraction, 
is thus obtained systematically in terms of atomic field correlation
functions.
The derivation leading to Eqs.~(\ref{eq4.3}) and (\ref{eq4.4}) does not 
depend on $\phi_{\rm b}$ being a bound state. The number of pairs of
atoms in any two body state is obtained in an analogous way. 
In Subsection \ref{sec2.5} we will apply an analogue of Eqs.~(\ref{eq4.3}) 
and (\ref{eq4.4}) to determine the number of atoms emitted from the 
condensate during the magnetic field pulse.

%
%
\subsection{Burst of atoms}
\label{sec2.5}
\begin{figure}[tb]
\begin{center}
\epsfig{file={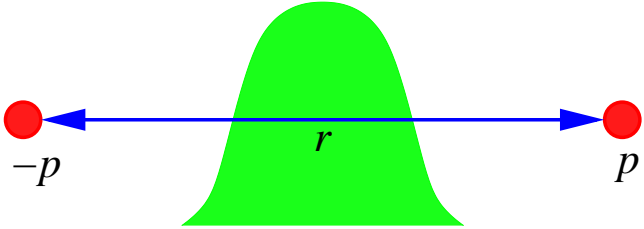},height=3cm,width=8.5cm,angle=0}
\vspace*{5mm}
\caption{\label{fig:expansion}Scheme of a ballistic expansion after a 
magnetic field pulse. Pairs of atoms are emitted from the condensate with 
one-particle momenta $\pv$ and $-\pv$. The center of mass of the pairs stays 
confined in the remnant condensate with a momentum spread determined by the 
spread of momenta in the condensate. At a sufficiently long time after the 
pulse a burst of relatively hot atoms can be detected outside the remnant 
condensate.}
\end{center}
\end{figure}

\noindent
In this subsection we show that the scattering part of the 
non condensed fraction on the right hand side of Eq.~(\ref{eq4.12}) 
determines the number of relatively hot atoms emitted in 
pairs from the condensate. To this end we consider a ballistic expansion 
of the gas at time $t_{\rm fin}$, i.e.~the trap is switched off and the 
magnetic field is held constant immediately at the end of a magnetic field 
pulse. This is illustrated schematically in Fig.~\ref{fig:expansion}.

A relatively 
hot fraction, if present, will expand much faster than the condensate and 
can be detected far outside the remnant condensate at a time $t$ sufficiently 
long after $t_{\rm fin}$. For a gas with $N$ atoms the observable for the 
number of pairs of atoms with a relative coordinate between $\rv$ and 
$\rv+d^3r$, in its first quantization form, is given by
$\sum_{i < j} |\rv_{ij}\ra\la\rv_{ij}|d^3r$,
where, as in Subsection \ref{sec2.3}, $i$ and $j$ indicate the pair of
atoms. In complete analogy to Subsection \ref{sec2.3} the mean number of 
pairs of atoms with a relative coordinate between $\rv$ and $\rv+d^3r$ 
becomes
\be
\label{eq4.14}
  n_{\rv}(t)d^3r
  = \intdd{R}|\Psi_{\rv}(\Rv,t)|^2 d^3r,
\ee
where
\bea
\label{eq4.15}
  &&\Psi_{\rv}(\Rv,t)
  \ =\ \frac{1}{\sqrt{2}}\lk[
     \Phi(\Rv,\rv,t)
    +\Psi(\Rv+\kfrac{\rv}{2},t)\Psi(\Rv-\kfrac{\rv}{2},t)\rk].
  \nonumber\\
\eea
%

At the relative distances $r$ under consideration, which exceed
by far the size of the remnant condensate, the second, factorized 
contribution to Eq.~(\ref{eq4.15}) is negligible. The energy spectrum of the
relatively hot atoms can be obtained from an expansion of $\Psi_{\rv}$ in
terms of the energy states in Eq.~(\ref{eq4.115}) that correspond to a release
of the atoms from the trap:
\bea
\nonumber
   \Psi_{\rv}(\Rv,t)
   =&& \frac{1}{\sqrt{2}}
    \sum_\nu\phi_{{\rm b}\nu}(\rv)\la\Rv,\phi_{{\rm b}\nu}|\Phi(t)\ra\\
   && + \frac{1}{\sqrt{2}}\intdd{p}
   \phi_{\pv}^{(+)}(\rv)\la\Rv,\phi_{\pv}^{(+)}|\Phi(t)\ra.
\label{eq4.16}
\eea
The molecular wave functions in Eq.~(\ref{eq4.16}) have decayed at the 
relevant distances $r$ that even exceed the extent of the remnant condensate. 
The corresponding molecular contribution to the right 
hand side of Eq.~(\ref{eq4.16}) is thus negligible. Taking into account that 
the scattering wave functions $\phi_{\pv}^{(+)}$ are energy eigenstates of a
pair of atoms after the pulse the remaining part of the amplitude in 
Eq.~(\ref{eq4.16}) becomes, after a short calculation using 
Eq.~(\ref{eq4.11}),
\bea
\nonumber
   \Psi_{\rv}(\Rv,t)
   =&&\frac{1}{\sqrt{2}}\intdd{p}
   \phi_{\pv}^{(+)}(\rv)\la\Rv,\phi_{\pv}^{(+)}|\Phi(t_{\rm fin})\ra\\
   &&\times
   e^{-i\frac{\pv^2}{m}(t-t_{\rm fin})/\hbar}.
\label{eq4.161}
\eea
For two identical atoms the relative kinetic energy $E_{\rm rel}$ and the
relative momentum $\pv$ are related through $E_{\rm rel}=\pv^2/m$.
The spectrum of the pairs of comparatively hot atoms, i.e.~the 
number of pairs of atoms with a relative energy in the interval 
$E_{\rm rel}\ldots E_{\rm rel}+dE_{\rm rel}$, is thus given by:
\bea
\nonumber
n(E_{\rm rel})dE_{\rm rel}
=&& \frac{1}{2}\sqrt{m}^3\sqrt{E_{\rm rel}}\ dE_{\rm rel}\\
&&\times  
\int d\Omega_{\pv}\intdd{R}|\Psi_{\pv}(\Rv)|^2,
\label{eq4.17}
\eea
where $\Psi_{\pv}(\Rv)=\la\Rv,\phi_{\pv}^{(+)}|\Phi(t_{\rm fin})\ra/\sqrt{2}$
is the amplitude on the right hand side of Eq.~(\ref{eq4.161}) 
and $d\Omega_{\pv}$ denotes the angular component of $d^3p$. The sum over 
all energy components of the spectrum yields 
\be
\int_0^\infty n(E_{\rm rel})dE_{\rm rel}=\frac{1}{2}\intdd{p}\intdd{R}
|\la\Rv,\phi_{\pv}^{(+)}|\Phi(t_{\rm fin})\ra|^2.
\label{eq4.171}
\ee

A comparison between Eqs.~(\ref{eq4.12}), (\ref{eq4.3}) and (\ref{eq4.171})
shows that the total non-condensed fraction of the gas consists of 
molecules and a burst of comparatively hot atoms emitted in pairs from the 
condensate with a time of flight spectrum of relative energies given by 
Eq.~(\ref{eq4.17}). Whether the non condensed fraction becomes significant 
depends on the time dependence of the magnetic field, i.e.~the way energy is
released to the gas. 
%
%
\section{Dynamics of the gas}
\label{sec3}
%
%
\subsection{Feshbach resonance and magnetic field pulse}
\label{sec1.1}
\begin{figure}[tb]
\begin{center}
\epsfig{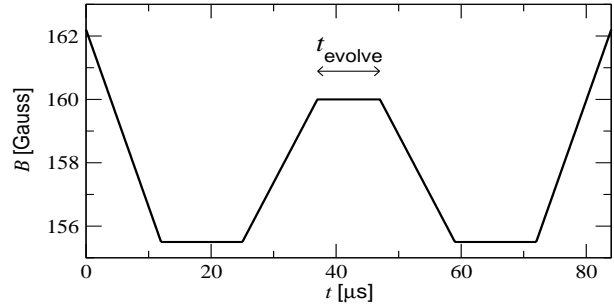}
\vspace*{5mm}
\caption{\label{fig1}Time dependence of the magnetic field.
The field varies linearly within the subsequent time
intervals. The numerical simulations in this article have been performed 
with the following pulse shape:
Fall and rise times: $12\,\mu$s; hold time at $B=155.5\,$G: $13\,\mu$s;
evolution time at $B=160\,$G: $t_{\mathrm evolve}=10...40\,\mu$s; initial and
final fields: $B=162.2\,$G.
}
\end{center}
\end{figure}

\noindent
In this Section we discuss the evolution of the gas when a specially designed  
homogeneous magnetic field pulse is applied to tune the inter-atomic 
interaction rapidly in the vicinity of a Feshbach resonance.
Motivated by the experiment of Donley et al.~\cite{Donley02}
we study the time variation of the magnetic field shown
in Fig.~\ref{fig1}.
The magnetic field varies linearly in time within the subsequent
time intervals.\\[2mm]
A Feshbach resonance occurs when the energy of a bound state of a closed
channel potential is tuned close to the dissociation threshold of the ground
state potential \cite{Feshbach}.
This tuning of the interaction in the inter-atomic motion takes advantage 
of the Zeeman effect in the electronic energy levels of the atoms.
If the closed channel bound state approaches the threshold from below,
the inter-atomic potential supports a shallow (metastable) 
$s$ wave bound state.
Around the resonance, a slight change in the energy difference of the
potentials thus leads to a large variation of the scattering length.
Neglecting the slow decay of the $s$ wave bound state, the scattering length 
depends on the magnetic field through the relation
\be
\label{eq1.1}
  a(B) = a_{\mathrm bg}\lk(1-\frac{\Delta B}{B-B_0}\rk),
\ee
where $\Delta B$ is the width of the Feshbach resonance and $B_0$ is the
resonant field.
We consider the resonance of $^{85}$Rb at $B_0=154.9\,$G, 
with $\Delta B=11.0\,$G
\cite{Roberts01}, which has been used in \cite{Donley02}.
For the background scattering length we use the value 
$a_{\mathrm bg}=-450\,a_{\mathrm Bohr}$
\cite{Donley02}, 
where $a_{\mathrm Bohr}$ is the 
Bohr-radius.\\
\begin{figure}[tb]
\begin{center}
\epsfig{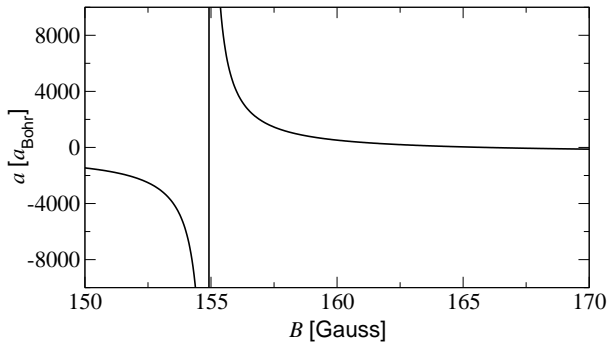}
\vspace*{5mm}
\caption{\label{fig2}The scattering length $a$ in units of the Bohr radius,
as a function of the magnetic field $B$, in the vicinity of the
Feshbach resonance. $a(B)$ is determined from Eq.~(\ref{eq1.1}), 
using $B_0=154.9\,$G, $\Delta B=11.0\,$G, and 
$a_{\mathrm bg}=-450\,a_{\mathrm Bohr}$.}
\end{center}
\end{figure}
Figure \ref{fig2} shows the scattering length as a function of the field
$B$ in the vicinity of the resonance for the values of $B$ that
are relevant in this article.
At the initial time $t_0$, the magnetic field $B=162.2\,$G implies a scattering
length of about $228\,a_{\mathrm Bohr}$ (cf.~Fig.~\ref{fig2}).
The interactions then vary according to the pulse shape in
Fig.~\ref{fig1}.
Similar to the experimental procedure \cite{Donley02} we will determine
the dynamic evolution of the gas for fixed time constants and field
strengths of the initial and final pulses, but for different evolution times
$t_{\mathrm evolve}$.

In the experiment \cite{Donley02} an adiabatic field variation followed
the pulse sequence. 
Finally the trap and the magnetic field were switched off, 
and the gas freely expanded before the number of atoms in the remnant 
condensate as well as a burst of relatively hot atoms were detected by 
absorption imaging. A series of measurements was performed for varying 
evolution times $t_{\mathrm evolve}$.
The number of atoms in each component showed an oscillatory dependence 
on $t_{\mathrm evolve}$ with the frequency corresponding to the 
energy of the shallow two body $s$ wave bound state in the evolution period.  
Moreover, a fraction of missing atoms was found oscillating at the same 
frequency. An interesting side result reported in  
\cite{Donley02} is that
the visibility of the oscillations depended sensitively on the 
presence of the initial and final ramp very close to the resonance
(at 155.5 G in Fig.~\ref{fig1}). In the next subsections we shall explain 
these observations.
 
\subsection{Coupling function}
\label{sec3.1}
\noindent 
We will study first the coupling function of the non-Markovian 
non-linear Schr\"odinger Eq.~(\ref{eq2.20}) 
for magnetic field variations as shown in Fig.~\ref{fig1}. 
The coupling function
$h(t,\tau)$, given in Eq.~(\ref{eq3.9}), reflects the binary dynamics 
that enters the description of the condensate through Eq.~(\ref{eq2.20}).   
We will discuss to which extent the binary dynamics can already explain
why the particular field pulse was needed to observe the oscillations
between the condensate and the non condensed fraction of the gas.   

\begin{figure}[tb]
\begin{center}
\epsfig{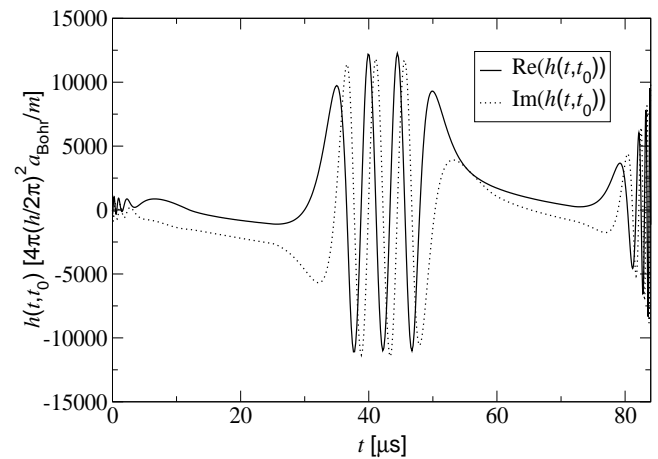}
\vspace*{5mm}
\caption{\label{fig3}The coupling function $h(t,\tau)$ as a function of $t$, 
describing the two-body dynamics driven by the magnetic field pulse in 
Fig.~\ref{fig1}, for
$\tau=t_0$, and an evolution time $t_{\mathrm evolve}=10\,\mu$s. 
}
\end{center}
\end{figure}
We have numerically determined the time dependence of $h(t,\tau)$ 
in the two dimensional 
plane $(t,\tau)$ for $t_0<t<t_{\rm fin}$ and $t_0<\tau<t$ using 
the methods described in Appendix \ref{appB}.
Figure \ref{fig3} shows $h(t,\tau)$ as a function of $t$, for $\tau=t_0$
and the time dependence of the magnetic field in Fig.~\ref{fig1}, with
$t_{\mathrm evolve}=10\,\mu$s.
In the evolution period between $t_3=37\,\mu$s and
$t_4=47\,\mu$s, $h(t,\tau)$ oscillates with the frequency 
$\nu_{\rm evolve}\cong 200\,$kHz. 
This particular oscillatory dependence is to be expected, as, according to 
Eq.~(\ref{eq3.9}), 
$h(t,\tau)$ involves the two body time development operator, 
$U_{\rm 2B}(t,\tau)$: In this period the binary
potential is stationary and supports a shallow $s$ wave bound state 
$\phi_{\rm b}^{\rm evolve}$. A spectral decomposition of $U_{\rm 2B}(t,\tau)$
shows that the contribution of this bound state to $h(t,\tau)$, within the
evolution period, is given by:
\bea
\nonumber
h(t,\tau)\cong && (2\pi\hbar)^3\la 0|V(t)|\phi_{\rm b}^{\rm evolve}\ra
\la\phi_{\rm b}^{\rm evolve}|U_{\rm 2B}(t_3,\tau)|0\ra\\
&&\times
\theta(t-\tau) e^{-i E_{\rm b}^{\rm evolve}(t-t_3)/\hbar},
\label{h_evolve}
\eea 
where $t_3$ is the initial time of the evolution period. The frequency of
the oscillations in the coupling function thus corresponds to the bound
state energy in the evolution period, 
i.e.~$\nu_{\rm evolve}=|E_{\rm b}^{\rm evolve}|/h$.
The amplitude and phase of these oscillations, however, depend on the time 
evolution before. 

\begin{figure}[tb]
\begin{center}
\epsfig{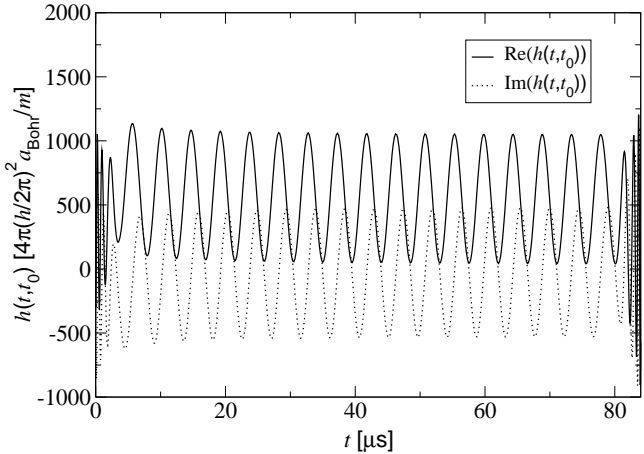}
\vspace*{5mm}
\caption{\label{fig4}The coupling function $h(t,\tau)$ as
a function of $t$, for $\tau=t_0$, describing the two-body dynamics for 
a trapezoidal pulse,
i.e.~the magnetic field pulse in Fig.~\ref{fig1} but without 
the initial and final ramps to $B=155.5$ G. 
The magnetic field is thus held 
constant at $B=160$ G from $t\protect\cong 4$ $\mu$s to 
$t\protect\cong 80$ $\mu$s.}
\end{center}
\end{figure}

The particularly large amplitude in Fig.~\ref{fig3} is 
achieved by the initial ramp close to the resonance at $B\cong 155\,$G in
Fig.~\ref{fig1}. To illustrate the role of the first ramp, Fig.~\ref{fig4} 
shows $h(t,\tau)$ as a function of $t$, at $\tau=t_0$, for a trapezoidal 
pulse.
Here, the trapezoidal pulse is chosen similar to Fig.~\ref{fig1} except that
the initial and final ramps to $B=155.5$ G are cut off, i.e.~the 
magnetic field is held constant at 160 G between $t\cong 4$ $\mu$s 
and $t\cong 80$ $\mu$s. 
As shown in Figs.~\ref{fig3} and \ref{fig4} the first ramp to $B=155.5$ G 
causes a pronounced enhancement of the amplitude by a factor of about 20 
for the optimized pulse in Fig.~\ref{fig1} as compared to the trapezoidal 
pulse. \\[5mm]  

\begin{figure}[tb]
\begin{center}
\epsfig{file={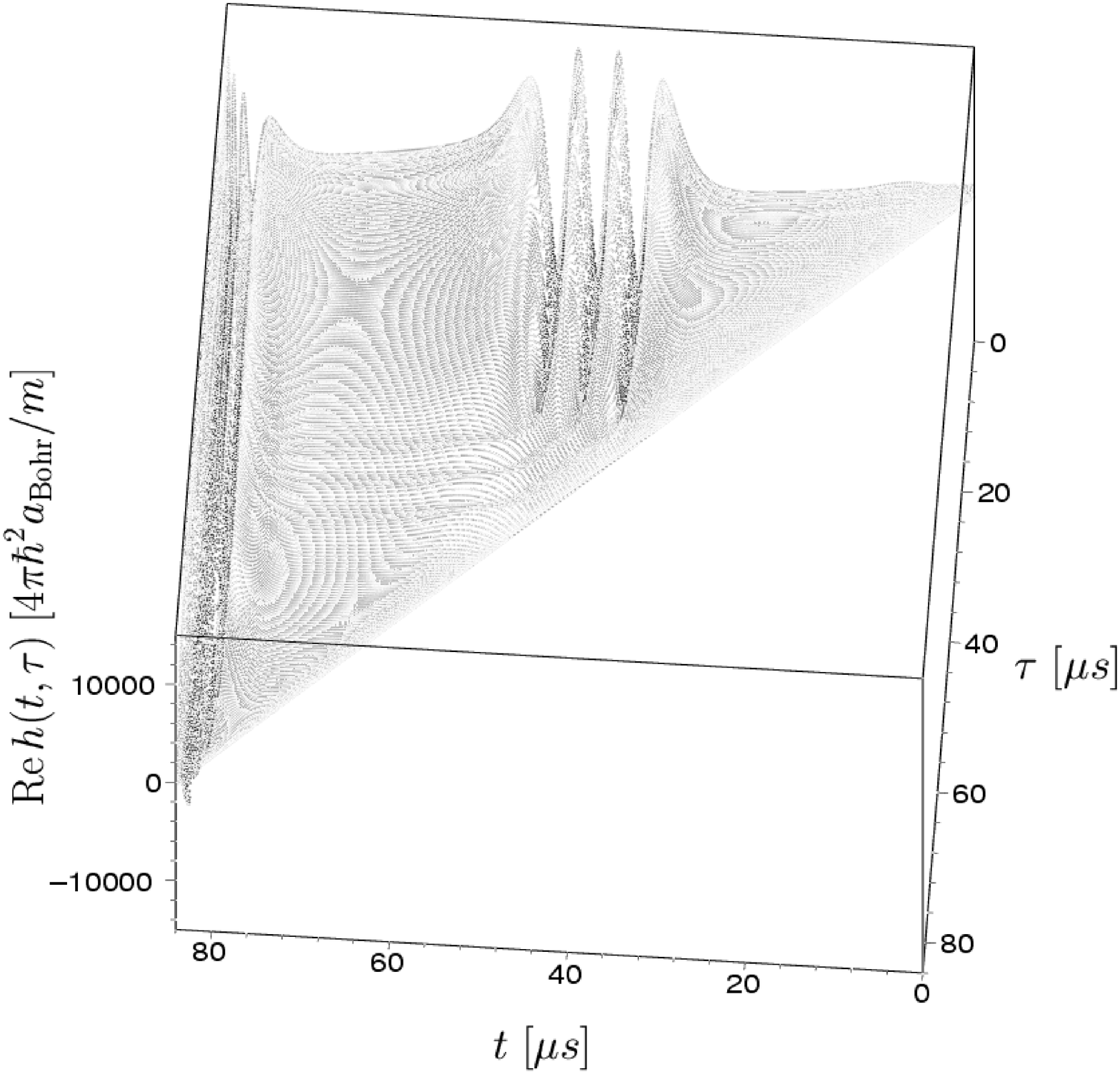},height=7cm,width=8cm,angle=0}
\vspace*{5mm}
\epsfig{file={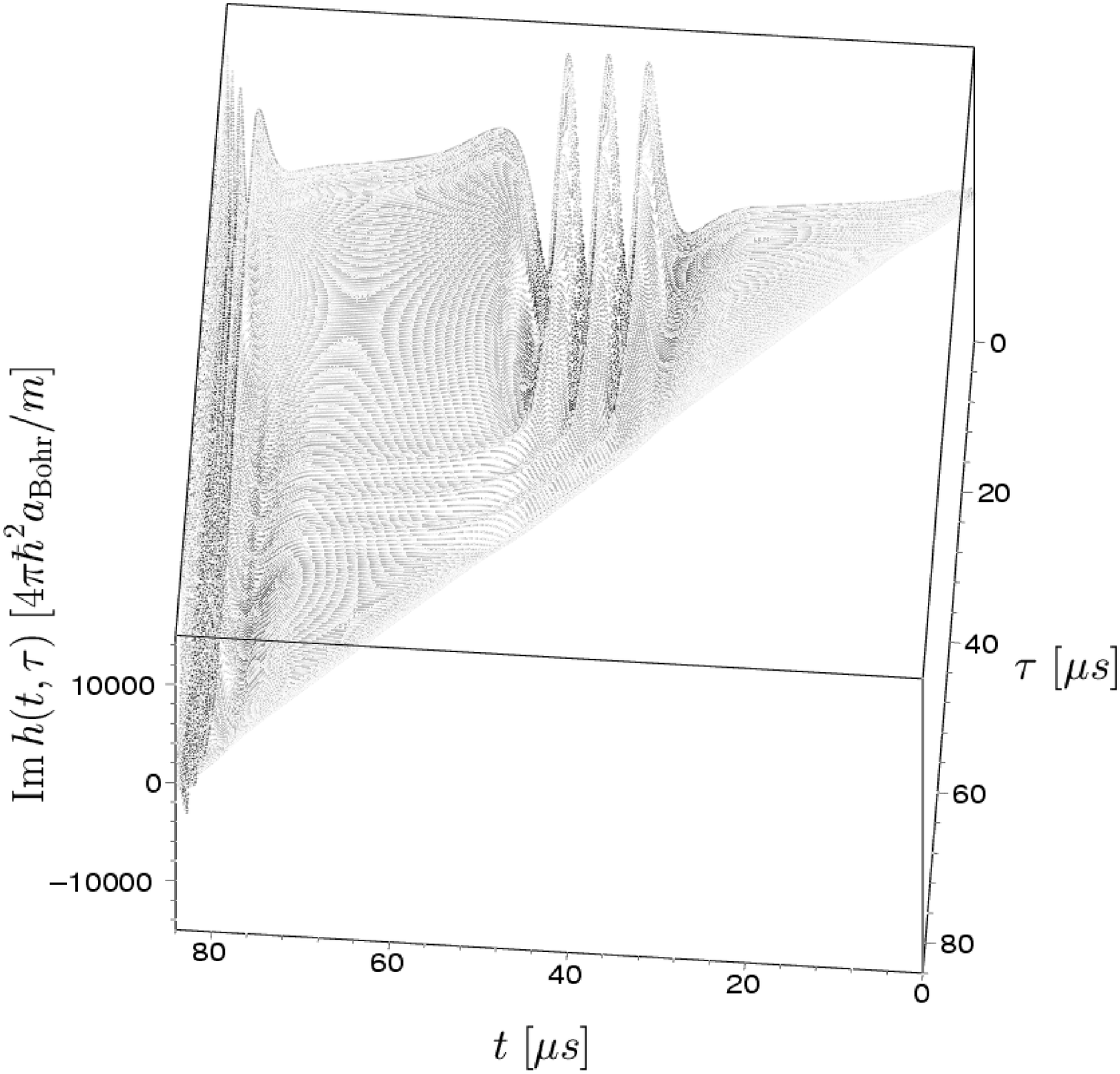},height=7cm,width=8cm,angle=0}
\vspace*{5mm}
\caption{\label{fig5}The coupling function $h(t,\tau)$ as a function of $t$ 
and $\tau$, for the same parameters as used in Fig.~\ref{fig3}.}
\end{center}
\end{figure}

Figure \ref{fig5} shows the real and imaginary part of 
$h(t,\tau)$ in the two-dimensional plane $(t,\tau)$, for the same parameters
as in Fig.~\ref{fig3}. The figure reveals that the amplitude of
the oscillations during the evolution period between $t=37$ $\mu$s and 
and $t=47$ $\mu$s rapidly decays in $\tau$. 
The phase of the oscillations, however, is largely independent of $\tau$.
A further analysis shows that these properties of the especially optimized 
pulse form in Fig.~\ref{fig1} assure the reappearance of the oscillation 
frequency of $h(t,\tau)$ in the nonlinear dynamics of the condensate 
described by Eq.~(\ref{eq2.20}).

%
%
\subsection{Homogeneous gas}
\label{sec4.1.1}
\noindent
The dynamics of a homogeneous condensate 
driven by a magnetic field pulse of the form in Fig.~\ref{fig1} already
exhibits all basic qualitative phenomena reported in 
\cite{Donley02}.  
We will therefore study the time evolution of the condensate as well as the 
final non-condensed fraction in detail for this idealized gas. 
Thereafter, we will discuss the corrections due to the presence of a trap in 
Subsection \ref{sec4.2}.
All physical quantities under consideration are determined by the 
nonlinear Schr\"odinger Eq.~(\ref{eq2.20}), driven by coupling functions
of the form of Fig.~\ref{fig5}, with a variable evolution time
$t_{\rm evolve}$.  
\begin{figure}[tb]
\begin{center}
\epsfig{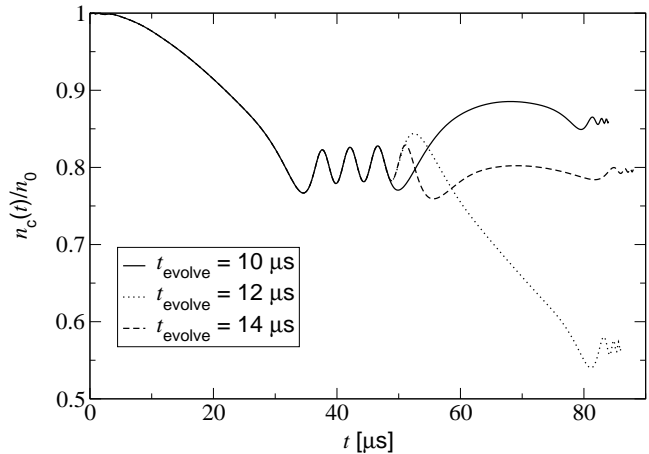}
\vspace*{5mm}
\caption{\label{fig6}The time dependence of the relative condensate fraction 
$n_{\rm c}(t)/n_0$ remaining
of the initial density of $n_0=3.9\times 10^{12}\,$cm$^{-3}$ for
three different evolution times $t_{\mathrm evolve}=10\,\,\mu$s, $12\,\mu$s,
and $14\,\mu$s. The sequence of magnetic field pulses is chosen as in 
Fig.~\ref{fig1} with a field strength in the evolution period of
$B_{\rm evolve}=160$ G.}
\end{center}
\end{figure}
Starting from a pure condensate with the density 
$n_0=3\times 10^{12}$ cm$^{-3}$, Fig.~\ref{fig6} shows the relative atomic 
condensate density $n_{\rm c}(t)/n_0$, as a function of $t$, for three 
different evolution times $t_{\mathrm evolve}=10\,\,\mu$s, $12\,\mu$s, 
and $14\,\mu$s. The initial conditions correspond roughly to the low 
density measurements in Ref.~\cite{Donley02}.
After an initial loss period during $t_0=0\le t\lsim 35\,\mu$s
the condensate density shows a distinct oscillatory behavior around 80\%
of the initial density.
The frequency of these oscillations very precisely matches the bound state
frequency in the evolution period, 
i.e.~$\nu_{\rm evolve}=|E_{\rm b}^{\rm evolve}|/h\cong 200\,$kHz.
After the evolution period, the second magnetic field pulse, which shifts the 
atoms in and out of the vicinity of the Feshbach resonance, causes the 
condensate fraction to develop to values between $55$\% and $85$\%.
The final fraction depends on the phase of the intermediate oscillations at
the end of the evolution period when the second resonant pulse starts. The
remnant condensate density at time $t_{\rm fin}$, immediately after the pulse
sequence, therefore, also oscillates as a function of $t_{\rm evolve}$.
While the first ramp to $B=155.5$ G in Fig.~\ref{fig1} drives the 
amplitude of the oscillations of $n_{\rm c}(t)$ in $t$ the second
ramp in Fig.~\ref{fig1} amplifies the visibility of the oscillations in
$n_{\rm c}(t_{\rm fin})$ as a function of $t_{\rm evolve}$.     
The fast oscillations of the function $h(t,\tau)$ in $t$ at the very
beginning and ending of the pulse sequence 
(see Figs.~\ref{fig3} and \ref{fig5})
have only a minor influence on the evolution of the condensate.\\[0.3cm] 
\begin{figure}[tb]
\begin{center}
\epsfig{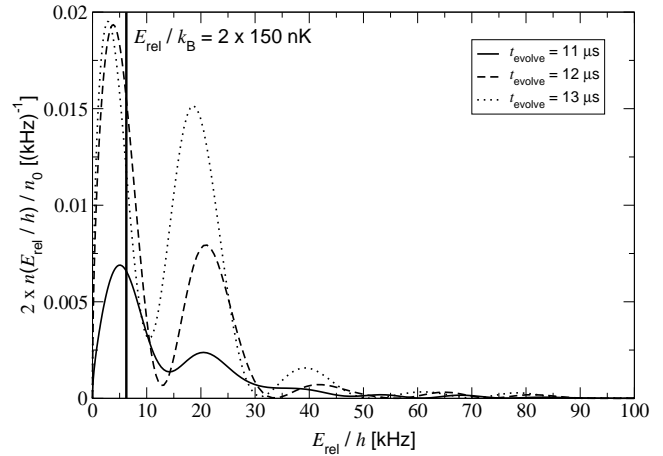}
\vspace*{5mm}
\caption{\label{fig:burst}The density of burst atoms 
($2\times n(E_{\rm rel})$ with $n(E_{\rm rel})$ given by 
Eq.~(\ref{eq4.17})) as a function of the relative energy $E_{\rm rel}$ 
in a uniform gas for three different evolution times 
$t_{\mathrm evolve}=11\,\,\mu$s, $12\,\mu$s, and $13\,\mu$s. The external 
parameters are chosen as in Fig.~\ref{fig6}. The atoms are emitted 
from the condensate in pairs with momenta $\pv$ and $-\pv$. 
The relative energy
$E_{\rm rel}=\pv^2/m$ is related to the energy of a single atom, 
$E_{\rm 1B}=\pv^2/2m$, through $E_{\rm 1B}=E_{\rm rel}/2$.  
The vertical line indicates the scale of the mean energies of burst
atoms reported in Ref.~\protect\cite{Donley02}.}
\end{center}
\end{figure}
The atomic mean field $\Psi$ determines the pair function through
Eq.~(\ref{eq4.11}) 
and, in turn, the molecular fraction in Eqs.~(\ref{eq4.3}) and (\ref{eq4.4})
as well as the energy spectrum of comparatively hot atoms, Eq.~(\ref{eq4.17}), 
after the pulse.
Figure \ref{fig:burst} shows the density of atoms emitted in pairs from the
condensate as a function of their relative energy for a uniform gas under
the conditions described in Fig.~\ref{fig6}. As the momentum spread 
of the center of mass motion of the pairs corresponds to the small spread 
of momenta in the atomic condensate the energy of a single atom in a pair is 
related to the relative energy through $E_{\rm 1B}=E_{\rm rel}/2$.  
The spectra exhibit a damped oscillatory dependence on the energy with a 
first, dominant maximum below $E_{\rm 1B}/k_{\rm B}=150$ nK. 
The time of 
flight energy spectra in Fig.~\ref{fig:burst}, as described in 
Subsection \ref{sec2.5}, do not correspond completely to the experimental
procedure in \cite{Donley02}. The presence of a trap should also modify their 
shape in a noticeable way. We expect, however, that Fig.~\ref{fig:burst} 
reflects the typical energy scales of the burst atoms in Ref.~\cite{Donley02}.
     
The experimental procedure did not allow for a direct 
detection of molecules. We thus identify the fraction of missing atoms 
reported in \cite{Donley02} as those atoms that are bound to dimer 
molecules after the pulse sequence. The total density of unbound atoms
is then given by the initial density $n_0$ minus twice the density of
dimer molecules in the homogeneous gas.
In the course of our studies we have determined
the remnant condensate as well as the final molecular fraction as a function
of the evolution time $t_{\rm evolve}$ from 10 to 40 $\mu$s in steps of 
1 $\mu$s. The total length of the pulses has thus been varied between 
$84\,\mu$s and $114\,\mu$s. 

\begin{figure}[tb]
\begin{center}
\epsfig{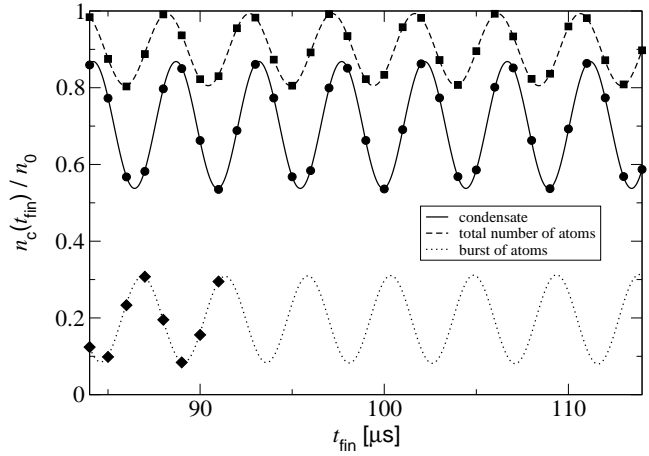}
\vspace*{5mm}
\caption{\label{fig7}The remaining fraction of condensate atoms, 
$n_{\rm c}(t_{\rm fin})$,
(solid line) together with the total density of unbound atoms (dashed line)
in a homogeneous gas, as a function of the final time $t_{\rm fin}$, 
elapsed after a magnetic field pulse of the form in Fig.~\ref{fig1}. All
densities are given relative to the initial density of
$n_0=3.9\times 10^{12}$ cm$^{-3}$. The external parameters are chosen as 
in Fig.~\ref{fig1}. The filled circles and squares correspond to direct
calculations of the remnant condensate and the molecular fraction. The
solid and dashed curves are interpolations with the sinusoidal fit functions 
proposed in Ref.~\protect\cite{Donley02}. The dotted line indicates the
``burst'' of relatively hot (unbound) atoms emitted in pairs from 
the condensate as determined from the remnant condensate and the total 
fraction of unbound atoms through number conservation 
(cf.~Eqs.~(\ref{eq4.113}) and (\ref{eq4.12})). The filled diamonds correspond 
to direct calculations of the ``burst'' fraction obtained from integration
of the spectra in Fig.~\ref{fig:burst} with respect to the energy.}
\end{center}
\end{figure}

The results are summarized in Fig.~\ref{fig7}.
The solid line is an interpolation of the data for the remnant condensate
density relative to the initial density of $n_0=3.9\times 10^{12}$ cm$^{-3}$
(filled circles) with the sinusoidal
fit function proposed by Donley et al.~\cite{Donley02}.
The frequency of the oscillations corresponds to the binding
energy $|E_{\rm b}^{\rm evolve}|/h=|E_{\rm b}(160\,$G$)|/h\cong 200\,$kHz.
In Fig.~\ref{fig7} the filled squares and their interpolation, i.e.~the 
uppermost curve, show the fraction of atoms which are not bound
to dimer molecules after the magnetic field pulse (Fig.~\ref{fig1}).
Number conservation allows to determine the density of the burst of relatively
hot atoms directly from the total density of unbound atoms and the
remnant condensate (cf.~Eqs.~(\ref{eq4.113}) and (\ref{eq4.12})).  
For this reason the dotted curve in Fig.~\ref{fig7}, termed 
``burst of atoms'', has been obtained by subtracting the solid 
curve from the dashed curve. Both the fraction of ``missing'' atoms and
the ``burst'' of atoms exhibit oscillations with the frequency 
$\nu_{\rm evolve}=|E_{\rm b}^{\rm evolve}|/h$ in $t_{\rm evolve}$ as 
well as the phase relation with respect to the remnant condensate   
reported in \cite{Donley02}.
%
%
\subsection{Trapped gas}
\label{sec4.2}
\noindent
The studies of the homogeneous gas in Subsection \ref{sec4.1.1} allowed to 
identify the three components of the gas as observed in \cite{Donley02}.
The purpose of this subsection is to study the influence of the trap and
the inhomogeneous local densities of the gas on the relative magnitudes of 
these components.\\
\begin{figure}[tb]
\begin{center}
\epsfig{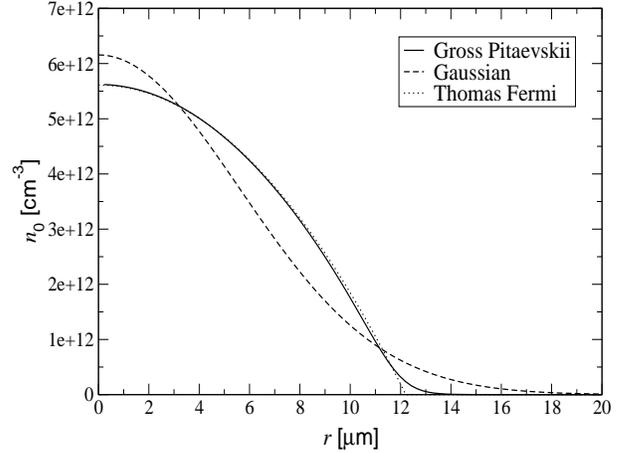}
\vspace*{5mm}
\caption{\label{fig8}The different initial states studied in a spherical 
harmonic trap with an oscillator length of about 
$l_{\mathrm ho}\protect\cong 3\,\mu$m,
i.e.~a trap frequency of $\omega_{\mathrm ho}\protect\cong 80\,$s$^{-1}$, 
corresponding to the geometric mean of the frequencies used in the 
experiment \protect\cite{Donley02} 
($\omega_{\mathrm ho}=\sqrt[3]{\omega_{\rm radial}^2\omega_{\rm axial}}$).
The atoms are exposed to a magnetic field of $B=162.2\,$G which implies
a scattering length of about 228 $a_{\rm Bohr}$.
Shown are the densities as functions of the radius corresponding to the 
exact ground state wave function given by the time independent 
Gross-Pitaevskii equation (solid line), and a Gaussian
(dashed line), which is the best least square fit to the
Thomas-Fermi approximation (dotted line). All densities correspond to
a number of 17100 condensed atoms.}
\end{center}
\end{figure}
We have performed these studies similar to Subsection \ref{sec4.1.1}
but in the presence of a trap. 
The trap has been idealized as a spherical harmonic oscillator potential 
$V_{\mathrm trap}(r)=\eh m\omega_{\mathrm ho}^2r^2$ with an oscillator 
frequency of about $\omega_{\mathrm ho}\cong 80$ s$^{-1}$ and a 
resulting oscillator length of about $l_{\mathrm ho}\cong 3\,\mu$m.
The trap parameters correspond to the geometric mean of the frequencies
in Ref.~\cite{Donley02}, 
i.e.~$\omega_{\mathrm ho}=\sqrt[3]{\omega_{\rm radial}^2\omega_{\rm axial}}$.
We have studied initially pure condensates of $N_0=17100$ atoms with
different local densities.
The different initial density profiles are shown in Fig.~\ref{fig8}.
The solid line is the density obtained from the exact solution to the time
independent Gross-Pitaevskii equation for 17100 atoms with a scattering 
length of about $a(162.2\,$G$)\cong 228\,a_{\mathrm Bohr}$.
This stationary condensate ground state exhibits a nearly perfect agreement
with its Thomas Fermi approximation. The second state we have studied is
a Gaussian (dashed line) which, for $17100$ atoms, is the best least
square fit to the Thomas-Fermi approximation (dotted line).

\begin{figure}[tb]
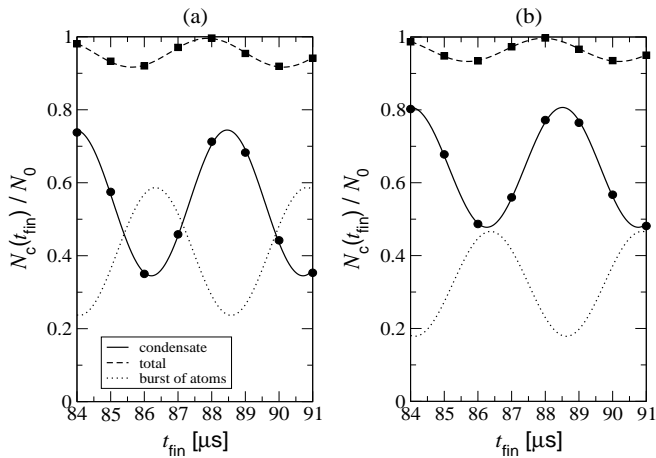

\begin{center}
\epsfig{file={fig11a.eps},height=6cm,width=4.2cm,angle=0}\ \
\epsfig{file={fig11b.eps},height=6cm,width=4.2cm,angle=0}
\vspace*{5mm}
\caption{\label{fig9}The remaining fraction of condensate atoms, 
$N_{\rm c}(t_{\rm fin})$,
(solid line), together with the molecular fraction (dashed line) and the 
burst of relatively hot unbound atoms (dotted line), 
as a function of the total time $t_{\rm fin}$ 
elapsed after the magnetic field pulse. The fractions are given relative
to the initial number of atoms, i.e.~$N_0=17100$.  
The calculations take into account the exact quantum dynamics of the trap for
the initial state given, in Fig.~\ref{fig8}, by the time independent 
Gross-Pitaevskii equation (a) and a Gaussian density distribution (b).}
\end{center}
\end{figure}
Figure \ref{fig9} shows the condensate, molecular and burst fractions
at the end of the pulse (Fig.~\ref{fig1}) for the exact solution of the 
time independent Gross-Pitaevskii equation (a) and the Gaussian (b). 
The evolution times range between $10\,\mu$s and $17\,\mu$s such that 
the total time of the pulse ranges between $84\,\mu$s and $91\,\mu$s.
The analysis of the trapped gas does not show any qualitative differences
from its analogue in a homogeneous gas in Fig.~\ref{fig7}. All components
oscillate with the same frequency 
$\nu_{\rm evolve}=|E_{\rm b}^{\rm evolve}|/h$ and exhibit phase shifts
similar to those in Fig.~\ref{fig7}.
Figure \ref{fig9} reveals, however, that the relative magnitude of the 
components depends sensitively on the local densities in the initial
inhomogeneous condensate.

\begin{figure}[tb]
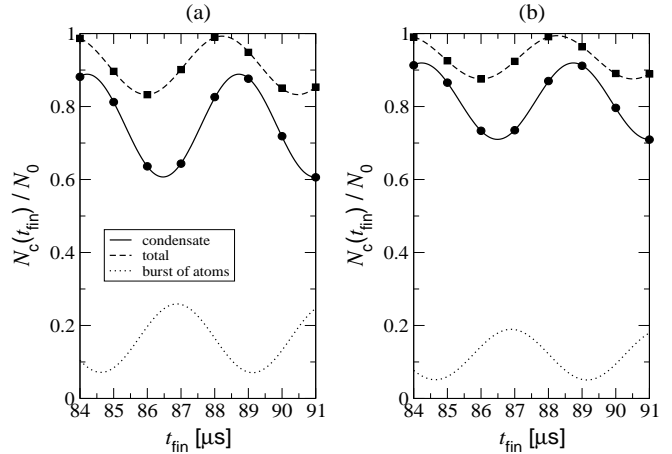

\begin{center}
\epsfig{file={fig12a.eps},height=6cm,width=4.2cm,angle=0}\ \
\epsfig{file={fig12b.eps},height=6cm,width=4.2cm,angle=0}
\vspace*{5mm}
\caption{\label{fig10}The remaining fraction of condensate atoms,
$N_{\rm c}(t_{\rm fin})$,
(solid line), together with the molecular fraction (dashed line) and burst of 
relatively hot unbound atoms (dotted line), as a function of the total 
time $t_{\rm fin}$ 
elapsed after the magnetic field pulse. The fractions are given relative
to the initial number of atoms, i.e.~$N_0=17100$.
The calculations are performed in the local density approximation for
the initial state given, in Fig.~\ref{fig8}, by the time independent 
Gross-Pitaevskii equation (a) and a Gaussian density distribution (b).}
\end{center}
\end{figure}

To study the role of the trap potential and the one body kinetic energy 
we have performed the same analysis as in Fig.~\ref{fig9} in the local 
density approximation which accounts only for the dynamics of the nonlinear 
Schr\"odinger Eq.~(\ref{eq2.20}) for a uniform gas but with
different densities, weighted according to the initial condensate wave
function. The approximation, however, neglects the trapping potential 
and the one body kinetic energy. 
Figure \ref{fig10} shows the different fractions of the gas under the same  
conditions as in Fig.~\ref{fig9}.
A comparison of Figs.~\ref{fig9} and \ref{fig10} reveals
no qualitative differences, but a quantitative dependence of all fractions
of the gas on the trap potential and the one body kinetic energy. The 
oscillation frequencies are the same but the mean values and amplitudes vary.
In the local density approximation the fraction of burst atoms results 
considerably smaller than in the exact calculation. A further analysis
of the time dependence of the condensate fraction, similar to 
Fig.~\ref{fig6}, shows that the predominant influence of the trap potential 
occurs during the initial ramp to $B=155.5$ G in Fig.~\ref{fig1}. 
In this period of the pulse sequence the scattering length becomes comparable
to the oscillator length
\cite{Tiesinga00}. This additional length scale is neither accounted 
for in the homogeneous gas nor in the local density approximation.
The results of this subsection show that a quantitative comparison of the
magnitude of the different fractions in the gas with the experiment 
\cite{Donley02}
should include the correct trap potential with the precise local densities
in the initial condensate as well as the precise time dependence of the
magnetic field pulse.

%
%
\subsection{Interpretation of the results}
\label{sec4.5}
\noindent
In the preceding sections we have analyzed the dynamics of an initially 
condensed Bose gas of $^{85}$Rb atoms exposed to a magnetic field pulse
of the form in Fig.~\ref{fig1}. We have identified the different fractions of 
unbound atoms in the gas as a remnant condensate and a burst of 
comparatively hot atoms with kinetic energies of about 
$E_{\rm 1B}/k_{\rm B}\gsim 100$ nK. We have further analyzed the dependence
of the relevant physical observables on the precise external conditions.
Our results strongly indicate that the fraction of missing atoms reported
in \cite{Donley02} corresponds to atoms bound to dimer molecules that could
not be detected. 

Donley et al.~\cite{Donley02} raised the question whether these molecules 
form a condensate. The analysis in Section \ref{sec2} provides a definite 
answer on the basis of the microscopic approach to the many body quantum 
dynamics: In Subsection \ref{sec2.5} we have predicted that the
molecular fraction would stay confined in the atomic condensate in a 
ballistic expansion immediately after the pulse while the burst fraction of
unbound pairs of atoms rapidly disperses. The expansion served as the
first experimental technique to prove the presence of a condensate. 

\begin{figure}[tb]
\begin{center}
\epsfig{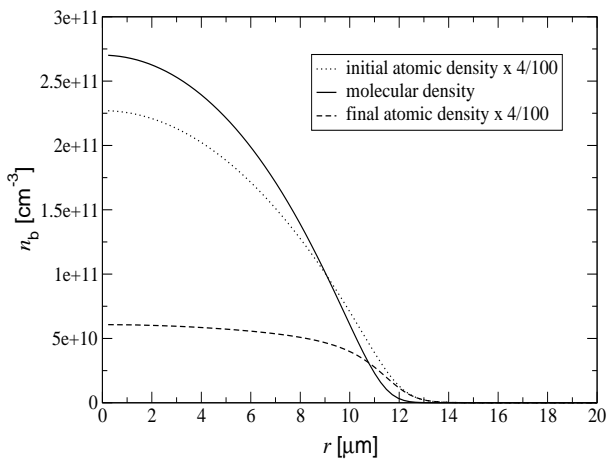}
\vspace*{5mm}
\caption{\label{fig11} The initial (dotted line) and final (dashed line) 
atomic condensate density, together with the molecular condensate 
density, $n_{\rm b}$, 
(solid line) for $t_{\mathrm evolve}=16\,\mu$s in a spherical trap. 
The curves correspond to the calculations in Fig.~\ref{fig9} (a). The atomic
densities are multiplied by a factor of 0.04.}
\end{center}
\end{figure}

The formation of a molecular condensate can be physically understood from the
nature of the external perturbation of the initial atomic condensate: 
The magnetic field pulse provides energy to form dimer molecules and correlated
pairs of burst atoms but no momentum to drive their centers of mass. 
A further analysis of Eqs.~(\ref{eq4.11}), (\ref{eq4.4}) and (\ref{eq4.161}) 
shows that, indeed, the molecules as well as the correlated pairs of burst 
atoms exhibit the same momentum spread in their centers of mass as the 
condensate atoms. Momentum is transferred only to the relative coordinate of 
the burst atoms such that the total momentum of the pairs remains 
negligibly small. A typical density of a molecular condensate in a spherical 
trap is shown in Fig.~\ref{fig11}.\\ 

As a great advantage of the microscopic dynamics approach, the molecular
condensate wave function, $\Psi_{\rm b}$ in Eq.~(\ref{eq4.4}), has been 
derived 
from the exact hierarchy of dynamic equations for correlation functions of 
field operators with the general Hamiltonian Eq.~(\ref{eq2.2}) using a 
systematic truncation scheme. Hence, the determination of the molecular 
condensate fraction does not rely upon any assumption on the existence of
a molecular order parameter. The analysis in this subsection is independent of
the precise trap geometry or the density profile of the initial state.  

%
%

\acknowledgements
\noindent
We would like to thank Eite Tiesinga, Eric Bolda, Paul Julienne, and
Bill Phillips for very inspiring and helpful discussions.
This work has been supported by the Alexander von Humboldt-Foundation 
(T.K., T.G.), the European Community under contract 
no.~HPMF-CT-1999-0023 (T.G.), and the United Kingdom EPSRC.

\begin{appendix}
%
%
\section{The microscopic quantum dynamics approach}
\label{appA}
\noindent
In this appendix we will derive the first order microscopic dynamics approach
to Bose condensed gases. The present derivation includes the explicit dynamics 
of the non-condensed fraction which is not considered in Ref.~\cite{Koehler02}.
%
%
\subsection{Dynamic equations for cumulants}
\label{secA.1}
\noindent
In general, the quantum state of a gas at time $t$ is described by a 
statistical operator $\rho(t)$ and the corresponding expectation value 
of an operator $\cO$ reads:
\be
\label{eq2.3}
  \la\cO\ra_t = \Tr\lk[\rho(t)\cO\rk].
\ee
Quantum expectation values of normal ordered products of field operators
are termed correlation functions.
The exact dynamics of the infinite hierarchy of correlation functions 
is given by
\be
\label{eq2.42}
  i\hbar\ptdnd{t}\la\psidg(\xv_n)\cdots\psi(\xv_1)\ra_t
  = \la[\psidg(\xv_n)\cdots\psi(\xv_1),H]\ra_t,
\ee
where $n$ denotes the number of field operators in the normal ordered 
product and $H$ is the quite general Hamiltonian in Eq.~(\ref{eq2.2}).
The set of Eqs.~(\ref{eq2.42}) is equivalent to the many body Schr\"odinger 
equation and determines all physical properties of a gas of atoms.
Due to the interaction term in Eq.~(\ref{eq2.2}) any finite subsystem of
Eqs.~(\ref{eq2.42}) does not close, and any attempt of an approximate solution
relies upon a consistent way of truncation.
A systematic truncation scheme, based on cumulants of correlation functions, 
has been proposed in \cite{Fricke96} for the dynamics of fermionic many body 
systems on short time scales. The extension of this work to interacting 
Bose gases
\cite{Koehler02}, 
with a modified truncation scheme, allows to describe the dynamics on the time 
scales that are relevant in this article.

The cumulants considered here are equivalent to the connected $n$-point
functions and are usually defined as derivatives of a generating functional
(cf., e.g., \cite{Weinberg}):
\bea
\label{eq2.5}
  &&\la\psidg(\xv_n)\cdots\psi(\xv_1)\ra^c=
\delndel{J(\xv_n)}\cdots\delndel{J^*(\xv_1)}
  \nonumber\\
  &&\quad\times\ \ln
    \lk.\lk\la
    e^{\intdd{x}[J^*(\xv)\psi(\xv)+J(\xv)\psidg(\xv)]}\rk\rangle
    \rk|_{J=J^*=0}.
\eea
The cumulants may also be derived recursively, and the first three orders 
read:
\bea
  \la\cO_1\ra
  &=& \la\cO_1\ra^c,
  \nonumber\\
  \la\cO_1\cO_2\ra
  &=& \la\cO_1\cO_2\ra^c+\la\cO_1\ra^c\la\cO_2\ra^c,
  \nonumber\\
  \la\cO_1\cO_2\cO_3\ra
  &=& \la\cO_1\cO_2\cO_3\ra^c+\la\cO_1\ra^c\la\cO_2\ra^c\la\cO_3\ra^c
  \nonumber\\
  &&+\ \la\cO_1\ra^c\la\cO_2\cO_3\ra^c+\la\cO_2\ra^c\la\cO_1\cO_3\ra^c
  \nonumber\\
  &&+\ \la\cO_3\ra^c\la\cO_1\cO_2\ra^c,
  \nonumber\\
  &\vdots&.
  \label{eq2.6}
\eea
For an ideal Bose gas, in the grand canonical thermal equilibrium, 
all cumulants containing more than two field operators vanish.
This is a consequence of Wick's theorem in statistical mechanics
according to which every number conserving normal ordered correlation
function can be expressed as a sum of products of all possible pair 
contractions conserving the operator ordering.
Moreover, as the expectation value of a single field vanishes, the second 
order cumulants become $\la\cO_1\cO_2\ra=\la\cO_1\cO_2\ra^c$. 
In accordance with
Eq.~(\ref{eq2.6}) the cumulants of an order higher than two vanish.
In an interacting gas the higher order cumulants thus provide a measure
for the deviation of the state of the gas from thermal equilibrium.

For the applications discussed in this article, with a condensate present 
in the gas, the relevant cumulants contain; 
the non number conserving condensate wave function 
$\Psi(\xv,t)=\la\psi(\xv)\ra_t^c$, 
the pair function $\Phi(\xv,\yv,t)=\la\psi(\yv)\psi(\xv)\ra_t^c$ and
the density matrix of the non condensed fraction 
$\Gamma(\xv,\yv,t)=\la\psi^\dagger(\yv)\psi(\xv)\ra_t^c$. 
The cumulant approach consists in transforming the exact hierarchy of
dynamic Eqs.~(\ref{eq2.42}) into an equivalent set of equations of
motion for cumulants.
The exact dynamic equations for the cumulants up to the second order 
are given explicitly in Ref.~\cite{Koehler02}.
The truncation scheme of Ref.~\cite{Koehler02} consists in retaining,
to the order of $n$, the exact dynamic equations for
cumulants up to the order of $n$ as well as the free time evolution of the
cumulants of the order of $n+1$ and $n+2$. The free time evolution of 
the cumulants of the order of $n+1$ and $n+2$ is obtained by neglecting, 
in their dynamic equations, all products of cumulants containing 
$n+3$ and $n+4$ field operators.

The first order microscopic dynamics approach ($n=1$) results in a closed 
nonlinear Schr\"odinger equation for the mean field $\Psi$ 
that allows us to describe the dynamics of a condensate close to but also far 
away from thermal equilibrium \cite{Koehler02}.
The first order dynamic equations that lead to the non-linear 
Schr\"odinger equation read \cite{Koehler02}:
\bea
   &&i\hbar\ptdnd{t}\Psi(\xv,t)
     = H_{\mathrm 1B}(\xv)\Psi(\xv,t)
   \nonumber\\
   &&\qquad+\ \intdd{y} V(\xv-\yv,t)\Psi^*(\yv,t)
   \nonumber\\
   &&\qquad\quad\times\Big[
     \Phi(\xv,\yv,t) + \Psi(\xv,t)\Psi(\yv,t)\Big],
   \label{eq2.8}\\
   &&i\hbar\ptdnd{t}\Phi(\xv_1,\xv_2,t)
     = H_{\mathrm 2B}(\xv_1,\xv_2)\Phi(\xv_1,\xv_2,t)
   \nonumber\\
   &&\qquad+\ V(\xv_1-\xv_2,t)\Psi(\xv_1,t)\Psi(\xv_2,t).
   \label{eq2.9}
\eea
Here the one and two-body Hamiltonians are denoted by
\bea
\nonumber
   H_{\mathrm 1B}(\xv)
   &=& -\hbar^2\nabv^2/2m+V_{\mathrm trap}(\xv),\\
   H_{\mathrm 2B}(\xv_1,\xv_2)
   &=& H_{\mathrm 1B}(\xv_1) + H_{\mathrm 1B}(\xv_2) + V(\xv_1-\xv_2,t).
   \nonumber\\
   \label{eq2.11}
\eea
Equations (\ref{eq2.8}) and (\ref{eq2.9}) include the loss of condensate
atoms into the non-condensed fraction as well as the back action of 
non condensed atoms on the condensate on time scales comparable to 
collisional durations. On longer time scales the non condensed fraction 
becomes dilute and its back action is neglected. 

The first order dynamics of the non-condensed fraction is 
determined through the conservation of the total number of atoms in the
gas:
\be
\label{eq2.10.1}
N=\intdd{x}\left[\Gamma(\xv,\xv,t)+|\Psi(\xv,t)|^2\right]
\ee
is a constant of motion.
The corresponding approximate dynamic equation for $\Gamma$ is then 
obtained from its exact counterpart in Ref.~\cite{Koehler02} through 
\bea
   &&i\hbar\ptdnd{t}\Gamma(\xv_1,\xv_2,t)
     = \bigg\{ H_{\mathrm 1B}(\xv_1)\,\Gamma(\xv_1,\xv_2,t)
   \nonumber\\
   &&\qquad+\ \intdd{y} V(\xv_1-\yv,t)\Phi^*(\yv,\xv_2,t)
   \nonumber\\
   &&\qquad\quad\times\Big[
     \Phi(\xv_1,\yv,t) + \Psi(\xv_1,t)\Psi(\yv,t)\Big]\bigg\}
   \nonumber\\
   &&\qquad-\ \{\xv_1\leftrightarrow\xv_2\}^*.
   \label{eq2.10}
\eea
The density matrix of the non-condensed fraction $\Gamma(\xv_1,\xv_2,t)$, 
as given in Eq.~(\ref{eq2.10}), 
is determined solely by the evolution of $\Psi$ from the
initial time $t_0$ up to the present time $t$. 

%
%
\subsection{First order dynamics}
\label{secA.2}
\noindent
In this subsection we will derive the non-linear Schr\"odinger 
Eq.~(\ref{eq2.20}) as well as Eqs.~(\ref{eq4.11}) and (\ref{eq4.112}) 
for the pair function and the density matrix of the non-condensed fraction, 
respectively. For the purpose of solving Eqs.~(\ref{eq2.9}) and 
(\ref{eq2.10}) formally in terms of the mean field $\Psi$ 
it is convenient to change the representation from the configuration space 
to the one body energy states of the trap potential $|\phi_i\ra$ 
or, for a homogeneous gas, into Fourier space. The corresponding single 
mode annihilation and creation operators obey the commutation relations
$[a_i,a^\dagger_j]=\delta_{ij}$ and the field operator becomes
$\psi(\xv)=\sum_i\phi_i(\xv)a_i$. In this new representation the cumulants 
up to the second order read: $\Psi_i(t)=\la a_i\ra^c_t$, 
$\Phi_{ij}(t)=\la a_ja_i\ra^c_t$, and 
$\Gamma_{ij}(t)=\la a_j^\dagger a_i\ra^c_t$. We will abbreviate
the trap states $|\phi_i\ra$ by $|i\ra$ in the following.

In the new representation Eq.~(\ref{eq2.9}) for the pair function
assumes the form
\bea
\nonumber
   &&i\hbar\ptdnd{t}\Phi_{ij}(t)
     =(E_i+E_j)\Phi_{ij}(t)\\
     \nonumber
   &&+\sum_{k_1,k_2}\la i,j\,|\,V(t)\,|\,k_1,k_2\ra
     \lk[\Phi_{k_1k_2}(t)+\Psi_{k_1}(t)\Psi_{k_2}(t)\rk],\\
   \label{eq2.12}
\eea
where $E_i$ is the eigenvalue of $H_\oneB$ with respect to the 
mode function $\phi_i(\xv)$.
Equation (\ref{eq2.12}) can be solved formally in terms of the
two body Green's function
\cite{Newton82}:
\bea
\nonumber
  &&\Phi_{ij}(t)
  = \sum_{k_1,k_2}\Big[
    \la i,j\,|U_{\rm trap}^\twoB(t,t_0)\,|\,k_1,k_2\ra\Phi_{k_1k_2}(t_0)\\
  \nonumber
  && + \int_{t_0}^t d\tau
    \la i,j\,|\,G_\twoB^{(+)}(t,\tau)V(\tau)\,|\,k_1,k_2\ra
  \Psi_{k_1}(\tau)\Psi_{k_2}(\tau)\Big].\\
  \label{eq2.14}
\eea
Here $G_\twoB^{(+)}$ is the retarded two-body Green's function,
\be
\label{eq2.15}
  \lk(ih\ptdnd{t}-H_\twoB(t)\rk)G_\twoB^{(+)}(t,\tau)
  = \delta(t-\tau),
\ee
which vanishes for $t<\tau$.
The retarded Green's function is related to the time development operator of
two trapped interacting atoms, given by the time ordered exponential
\be
\label{eq2.15a}
U_{\rm trap}^\twoB(t,\tau)=\cT\exp\left[
-\frac{i}{\hbar}\int_{\tau}^tdt'\,H_\twoB(t')\right], 
\ee
through
\be
\label{eq2.16}
  G_\twoB^{(+)}(t,\tau)=\frac{1}{i\hbar}
\theta(t-\tau)U_{\rm trap}^\twoB(t,\tau),
\ee
where $\theta(t-\tau)$ is the step function that yields unity for
$t>\tau$ and vanishes elsewhere.
In all applications in this article the gas is a dilute condensate at 
the initial time $t_0$. The
initial pair function $\Phi_{k_1k_2}(t_0)$ on the right hand side of 
Eq.~(\ref{eq2.14}) can then be neglected 
\cite{Koehler02} and Eq.~(\ref{eq2.14}) thus yields 
Eq.~(\ref{eq4.11}) in position space.

Inserting Eq.~(\ref{eq2.14}) into Eq.~(\ref{eq2.8}) for the mean field
leads to the closed non-linear Schr\"odinger equation
\bea
   &&i\hbar\ptdnd{t}\Psi_{i}(t)
     = E_i\Psi_{i}(t)
   \nonumber\\
   &&\qquad+\ \sum_{k_1,k_2,k_3}\int_{t_0}^\infty d\tau\,
     \la i,k_3|\,T_\twoB^{(+)}(t,\tau)\,|\,k_1,k_2\ra
   \nonumber\\
   &&\qquad\qquad\qquad\times\ \Psi_{k_1}(\tau)\Psi_{k_2}(\tau)\Psi^*_{k_3}(t),
   \label{eq2.17}
\eea
where $T_\twoB^{(+)}$ denotes the retarded two-body transition matrix in
the time domain:
\be
\label{eq2.18}
   T_\twoB^{(+)}(t,\tau)
   = V(t)\delta(t-\tau)+V(t)G_\twoB^{(+)}(t,\tau)V(\tau).
\ee
As in all applications in this article the trap potential is slowly varying
on the spatial scale determined by the range of the binary interaction 
$V(t)$ the thermodynamic limit in the relative motion of two atoms 
can be performed in the collision term in Eq.~(\ref{eq2.17}) \cite{Koehler02}. 
The coupling function in Eq.~(\ref{eq3.9}) thus involves the time development
operator of the relative motion of two atoms in free space, denoted by
$U_{\rm 2B}(t,\tau)$. Transformed back to the position space  
Eq.~(\ref{eq2.17}) then yields the nonlinear Schr\"odinger Eq.~(\ref{eq2.20}).

In the first order microscopic dynamics approach 
the pair function, as given by Eq.~(\ref{eq2.14}) with $\Phi_{k_1k_2}(t_0)=0$, 
determines the density matrix of the non condensed fraction:
Differentiation with respect to the time $t$ shows that 
\be
\label{eq2.19}
\Gamma_{ij}(t)=\sum_k \Phi_{ik}(t)\Phi^*_{jk}(t)
\ee
is the solution of Eq.~(\ref{eq2.10}) for an initial pure condensate,
i.e.~$\Gamma_{ij}(t_0)=0$, which derives Eq.~(\ref{eq4.112}). 
The density matrix of the
non-condensed fraction thus assumes the form of a partial trace over 
one coordinate of a two body pure state. The corresponding occupation
numbers, i.e.~the diagonal elements 
\be
\label{eq2.19a}
\Gamma_{ii}(t)=\sum_k |\Phi_{ik}(t)|^2,
\ee
are positive, independent of the specific choice of the basis set.

As shown in Subsections \ref{sec2.3} and \ref{sec2.5} the non-condensed 
fraction consists of a molecular part and a burst of atoms 
emitted in pairs from the condensate with a comparatively fast relative 
motion. This separation corresponds to a ballistic expansion of a gas
that is released from a trap.
The density of molecules in the bound state $\phi_{\rm b}$ is described by the 
mean field $\Psi_{\rm b}$ in Eq.~(\ref{eq4.4}). In analogy to the collision 
term of the non-linear Schr\"odinger Eq.~(\ref{eq2.17}) the molecular mean 
field can be expressed in terms of the atomic condensate wave function $\Psi$: 
\be
\label{eq4.5}
  \Psi_{\rm b}(\Rv,t)
  = -\frac{1}{\sqrt{2}}\int_{t_0}^\infty d\tau\,
     \Psi^2(\Rv,\tau)\ptdnd{\tau}h_{\rm b}(t,\tau).
\ee
The corresponding coupling function, $h_{\rm b}(t,\tau)$,
involves the overlap of the molecular bound state wave function 
$\phi_{\rm b}$ and the two body time development operator which, 
in all applications in this article, is excellently approximated by 
the thermodynamic limit:
\be
\label{eq4.6}
  h_{\rm b}(t,\tau)
  =(2\pi\hbar)^{3/2}\la\phi_{\rm b}\,|\,U_\twoB(t,\tau)\,|\,0\ra\theta(t-\tau).
\ee

The energy spectrum of the burst atoms in Eq.~(\ref{eq4.17}) involves an
amplitude $\Psi_{\pv}(\Rv)$, similar to Eq.~(\ref{eq4.5}), except that
the bound state $\phi_{\rm b}$ is replaced by the stationary scattering state
$\phi_{\pv}^{(+)}$ which is associated with the relative momentum $\pv$,
i.e.
\bea
\nonumber
\Psi_{\pv}(\Rv)=&&-\frac{1}{\sqrt{2}}
\int_{t_0}^{t_{\rm fin}} d\tau \int d^3R'd^3r'\\
\nonumber
&&\times\Psi(\Rv'+\rv'/2,\tau)\Psi(\Rv'-\rv'/2,\tau)\\
&&\times\frac{\partial}{\partial \tau}\la\Rv,\phi_{\pv}^{(+)}|
U_{\rm trap}^{\rm 2B}(t_{\rm fin},\tau)|\Rv',\rv'\ra.
\label{eq4.7}
\eea
Here, $t_{\rm fin}$ is the final time immediately after the pulse in 
Fig.~\ref{fig1}.
As $\phi_{\pv}^{(+)}(\rv)$ is not confined in space (see Appendix \ref{appB})
the coupling function corresponding to $\Psi_{\pv}(\Rv)$ should explicitly 
account for the discrete nature of the trap states also in the relative 
motion of two atoms. In Section \ref{sec3} we have determined the energy
spectrum of the burst atoms for a homogeneous gas, i.e.~in the absence of
a trap potential. The coupling function
of $\Psi_{\pv}$ then becomes similar to $h_{\rm b}(t,\tau)$ in 
Eq.~(\ref{eq4.6}). 

%
%
\section{Two-body dynamics}
\label{appB}
\noindent
In Appendix \ref{appA} we have formulated the many body dynamics of
a condensed gas in terms of the unitary time evolution operator of two atoms
interacting through their inter-atomic potential.    
In this appendix we provide a practical approach to determine the relevant 
low energy time evolution of two $^{85}$Rb atoms that serves as an input to 
the microscopic dynamic description of a partially condensed gas  
exposed to the time dependent magnetic field discussed in 
Section \ref{sec3}. 
The approach takes advantage of the fact that in all applications in this
article the binary interaction of $^{85}$Rb is dominated by the presence
of a shallow $s$ wave bound state.
%
%
\subsection{Resonance enhanced scattering}
\label{secB.0}
\noindent
In ultra-cold dilute gases the energies of two colliding atoms are usually 
sufficiently small for the differential cross sections to become isotropic.
In accordance with effective range theory
\cite{Newton82}
the $s$ wave scattering amplitude can then be expanded as 
\be
\label{eqB01}
f_0(k)=-a+ia(ka)+{\cal O}(k^2)=\frac{-a}{1+ika}+{\cal O}(k^2),
\ee
where $k$ is the wave number that is related to the relative momentum 
of two colliding atoms through $p=\hbar k$. Equation (\ref{eqB01}) 
effectively provides an expansion in terms of $kl$, where the first
dominant length scale $l$ is given by the $s$ wave scattering length $a$.
The next correction term involves the effective range of the
binary potential $V$
\cite{Newton82}, 
denoted as $r_{\rm eff}$ in the following. In general, both $a$ and 
$r_{\rm eff}$ depend sensitively on the detailed shape of $V$.

When the binary potential supports a shallow $s$ wave bound state the
scattering length is positive and may by far exceed all the other length scales
set by $V$. This situation is sometimes referred to as a zero energy resonance
\cite{Newton82}. The scattering amplitude is then given by $-a/(1+ika)$, as  
obtained from the right hand side of Eq.~(\ref{eqB01}), which corresponds
to the contact potential
\cite{Dalibard99}. Extending the collision energies $p^2/m$ into
the complex plane Eq.~(\ref{eqB01}) yields the $T$ matrix 
\cite{Newton82}
of the contact potential which assumes the separable, i.e.~factorized, form:
\be
\label{eqB02}
T_{\rm 2B}(z)=\frac{|\chi\ra\xi\la\chi|}{1+i\sqrt{mz/\hbar^2}a},
\ee
where $z=p^2/m+i\varepsilon$ is a complex energy variable and the complex
square root is chosen with a positive imaginary part. The wave function
$|\chi\ra$ and the amplitude $\xi$ are obtained as 
$\la \rv|\chi\ra=\delta(\rv)$ and $\xi=4\pi\hbar^2 a/m$, respectively.
The $T$ matrix determines all eigenstates of the two body Hamiltonian. 
The pole on the right hand side of Eq.~(\ref{eqB02}) indicates that
the contact potential with a positive scattering length supports a single
$s$ wave bound state with the binding energy 
\be
\label{eqB03}
E_{\rm b}=-\hbar^2/ma^2.
\ee

The separable form of Eq.~(\ref{eqB02}) is quite general whenever the 
$T$ matrix is dominated by the pole of a shallow $s$ wave bound state 
$\phi_{\rm b}$.
A spectral decomposition of the two body Hamiltonian then implies that 
at low collision energies the $T$ matrix is well approximated by
\cite{Newton82}
\be
\label{eqB04}
T_{\rm 2B}(z)=\frac{|\chi\ra\xi\la\chi|}{1-\xi\la\chi|G_0(z)|\chi\ra},
\ee
where 
\bea
\nonumber
|\chi\ra&=&V|\phi_{\rm b}\ra,\\ 
\xi&=&1/\la\phi_{\rm b}|V|\phi_{\rm b}\ra
\label{eqB05}
\eea
and $G_0(z)=(z+\hbar^2\Delta/m)^{-1}$ is the free energy dependent Green's
function of the relative motion of two atoms. In few body scattering theory
Eq.~(\ref{eqB05}) is usually referred to as the unitary pole approximation
\cite{Gloeckle83}.
A further analysis shows
that the $T$ matrix in Eq.~(\ref{eqB04}) has a pole at the exact bound state
energy of the potential $V$. The wave function $\chi(r)$ accounts for the 
spatial extent of $V$.

\begin{figure}[tb]
\begin{center}
\epsfig{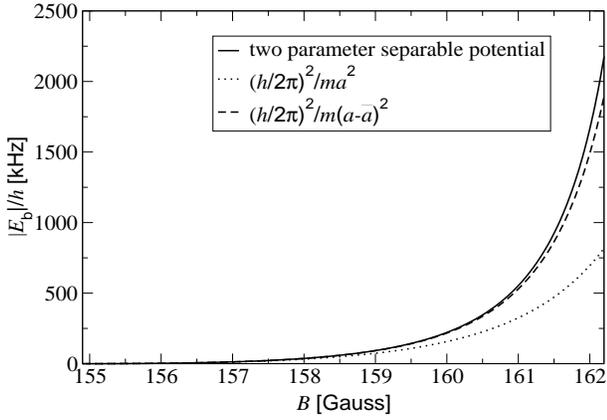}
\vspace*{5mm}
\caption{\label{fig14}
The energy of the uppermost $s$ wave bound state below threshold of two
$^{85}$Rb atoms, $E_{\rm b}$, as a function of the magnetic field strength $B$.
The dotted line shows Eq.~(\ref{eqB03}) which corresponds to a
zero energy resonance, with $a(B)$ determined from Eq.~(\ref{eq1.1}), 
using $B_0=154.9\,$G, $\Delta B=11.0\,$G, and 
$a_{\mathrm bg}=-450\,a_{\mathrm Bohr}$.
The dashed line shows Eq.~(\ref{eqB06}) which accounts for the van der
Waals interaction using $C_6=4660$ a.u. \protect\cite{Roberts01}. The 
dashed curve can be compared directly with a recent exact coupled channels 
scattering calculation \protect\cite{KokkelmansData}.
The solid line is obtained from the binding energies of two parameter 
separable potentials with the rounded parameter 
$\eta=5000 \times ma_{\rm Bohr}^2/\hbar$ and the
scattering length chosen in accordance with Eq.~(\ref{eq1.1}).}
\end{center}
\end{figure}

The long range behavior of inter-atomic potentials is determined by the van 
der Waals dispersion interaction $V_{\rm vdW}(r)=-C_6/r^6$. The spatial
extent of $V$ is then characterized by the van der Waals length 
$l_{\rm vdW}=(mC_6/\hbar^2)^{1/4}$. As shown by Gribakin and Flambaum
\cite{Gribakin93}
the next correction to the binding energy for
an inter-atomic potential modifies Eq.~(\ref{eqB03}) to
\cite{PaulJulienne}:
\be
\label{eqB06}
E_{\rm b}=-\hbar^2/m(a-\bar{a})^2.
\ee
Here $\bar{a}$ is the mean scattering length given in terms of the
$\Gamma$ function through
\be
\label{eqB07}
\bar{a}=\frac{1}{2\sqrt{2}} l_{\rm vdW} \frac{\Gamma(3/4)}{\Gamma(5/4)}. 
\ee
Figure \ref{fig14} illustrates the dependence of the bound state energy
of the shallow $s$ wave bound state of the $^{85}$Rb pair interaction on the
magnetic field $B$. The binding energies obtained from Eq.~(\ref{eqB06}) 
are sufficiently accurate to match a recent exact coupled channels 
scattering calculation \protect\cite{KokkelmansData}.
Although the $^{85}$Rb dimer, in the vicinity of the Feshbach resonance, 
is particularly weakly bound in comparison to usual molecular ground states 
Fig.~\ref{fig14} exhibits a pronounced difference between Eqs.~(\ref{eqB03})
and (\ref{eqB06}). This deviation from the zero energy resonance situation
is related to the large van der Waals length of about 
$l_{\rm vdW}=164\ a_{\rm Bohr}$.\\[0.3cm] 
%
%
\subsection{The separable potential approach}
\label{secB.1}
\noindent
Direct insertion shows that Eq.~(\ref{eqB04}) exactly solves the 
Lipp\-mann-Schwinger equation
\cite{Newton82} 
\be
\label{eqB11}
T_{\rm 2B}(z)=V+VG_0(z)T_{\rm 2B}(z)
\ee
as long as the actual potential $V$ is replaced by the separable
potential  
\be
\label{eqB12}
  V_{\mathrm sep}
  = |\chi\ra\xi\la\chi|.
\ee
Since the pioneering work of Lovelace
\cite{Lovelace64}
separable expansions of potentials
\cite{Belyaev90}
have played an important role in nuclear few body physics, 
as they provide a systematic approach to solve two body scattering problems 
analytically in a limited range of collision energies. In this subsection
we shall determine a separable potential of the form of Eq.~(\ref{eqB12})
that accurately describes the dynamics of two $^{85}$Rb atoms
in the relevant range of magnetic fields and collision energies. 

The unitary pole approximation in Eq.~(\ref{eqB05}) is obtained from
spectral properties of the two body Hamiltonian and thus applies to
inter-atomic potentials
\cite{HeKoe}. 
Equation (\ref{eqB05}) reproduces the exact bound state energy of 
$V$ but the scattering length is only 
approximate. A further improvement can be achieved by choosing $|\chi\ra$
and $\xi$ in such a way that the separable potential in Eq.~(\ref{eqB12})
matches both the energy of the last $s$ wave bound state and the scattering 
length of $V$ at the actual magnetic field. At the low collision momenta 
under consideration the corresponding plane wave states do not resolve
the functional form of the wave function $\chi(r)$.
The specific form of $\chi(r)$ is thus not relevant as long as 
$\chi(r)$ decays in space on the length scale set by the van der Waals length.
We have chosen a Gaussian form which in momentum space is given by 
\be
\label{eqB13}
  \chi(p) = \la\pv\,|\,\chi\ra = (2\pi\hbar)^{-3/2}e^{-\eta p^2/2m\hbar}.
\ee
In position space $|\chi\ra$ is of the form 
$\chi(r)\propto\exp(-r^2/2\sigma^2)$ with a range parameter
$\sigma=\sqrt{\hbar\eta/m}$. 
The separable potential is then parametrized by the two constants $\xi$ and 
$\eta$. These parameters have to be determined at each magnetic field through 
matching the binding energy and scattering length of the separable potential
to the values of $E_{\rm b}$ and $a$ of the actual interaction $V$.
The bound state energy of the separable potential is the real
energy $z=E_{\em b}$ at the pole of the $T$ matrix in Eq.~(\ref{eqB04}), 
i.e.~$E_{\rm b}$ is determined through
\be
\label{eqB14}
  1-\xi\la\chi\,|G_0(E_{\rm b})|\,\chi\ra  = 0.
\ee
The scattering length is obtained from the zero energy limit of the
$T$ matrix as
\bea
  a &=& (2\pi\hbar)^3\frac{m}{4\pi\hbar^2}\la0\,|\,T_{\rm 2B}(0)\,|\,0\ra
  \nonumber\\
  &=& \frac{m}{4\pi\hbar^2}
  \frac{(2\pi\hbar)^3|\la0\,|\,\chi\ra|^2}
       {1/\xi-\la\chi\,|G_0(0)|\,\chi\ra}.
  \label{eqB15}
\eea
It turns out that the optimized parameter $\eta$ is independent of $B$ with 
the corresponding range parameter $\sigma=\sqrt{\hbar\eta/m}$ 
roughly given by $l_{\rm vdW}/2$.
In the applications in this article we have used the rounded value of
$\eta=5000\times ma_{\rm Bohr}^2/\hbar$ and determined $\xi(B)$ in such a way 
that the separable potential matches exactly the dependence of the
scattering length on the magnetic field in Eq.~(\ref{eq1.1}). 
The resulting dependence of the binding energy of the separable potential on 
the magnetic field $B$ is shown in Fig.~\ref{fig14}.  
The separable potential approach, as proposed in this subsection, does not 
depend upon the accuracy of model potentials for all scattering channels 
and, instead, describes the complete low energy collision dynamics in terms 
of the scattering length $a$ and the van der Waals dispersion coefficient 
$C_6$. Both parameters of the actual binary potential are accessible to 
experiment
\cite{Roberts98,Roberts01}.  

To illustrate the degree of accuracy of the proposed approach we shall provide 
a comparison of the low energy scattering properties obtained from the 
separable potential with the exact solution of the Schr\"odinger equation
for a well known inter-atomic interaction. We have chosen the scattering of 
two ground state $^4$He atoms for this comparison as the binary potential 
supports a single shallow $s$ wave bound state and, within several decades of 
intensive study, all properties of the interaction have been determined very 
accurately, to a large extent, from first principle calculations
\cite{Tang95}.
The scattering length and the binding energy have been determined recently
from experiment in Ref.~\cite{Grisenti00}.

\begin{figure}[tb]
\begin{center}
\epsfig{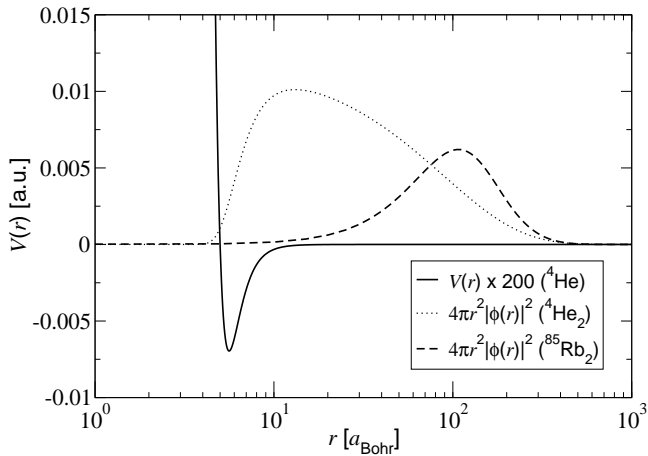}
\vspace*{5mm}
\caption{\label{fig13}The $^4$He binary (TTY) potential
\protect\cite{Tang95}, and the radial probability densities corresponding 
to the shallow $s$ wave bound
states of $^4$He as well as $^{85}$Rb, as obtained from the 
separable potential, at a magnetic field strength of $162.2\,$G. The
radius is given on a logarithmic scale.}
\end{center}
\end{figure}

\begin{figure}[tb]
\begin{center}
\epsfig{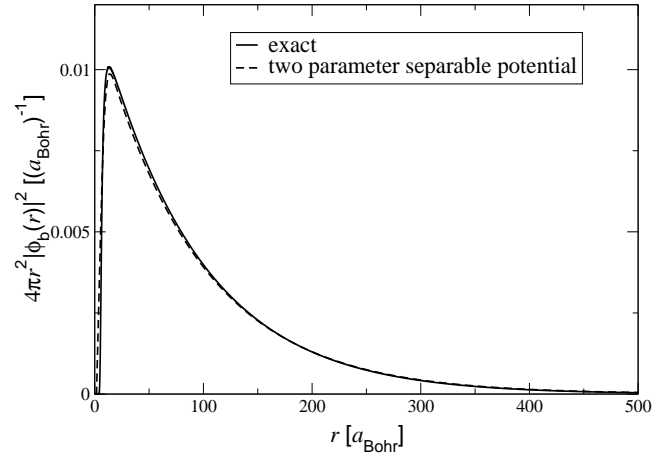}
\vspace*{5mm}
\caption{\label{figHe2sep}The $^4$He radial probability density 
corresponding to the $s$ wave bound state of $^4$He as obtained from
the TTY potential
\protect\cite{Tang95} (solid line) and the separable potential approach 
(dashed line).}
\end{center}
\end{figure}

We have determined the separable potential, as given through 
Eqs.~(\ref{eqB12}) and (\ref{eqB13}), that corresponds to the
Tang, Toennies and Yiu (TTY) $^4$He interaction 
\cite{Tang95} with $a=188\ a_{\rm Bohr}$ and $C_6=1.461$ a.u..
For $^4$He the ratio of the scattering length and the effective range,
$a/r_{\rm eff}$ is of the order of magnitude of 14. 
Figure \ref{fig13} shows the exact radial probability distribution of the
bound state wave function $^4$He$_2$ as well as the shallow $s$ wave 
bound state of $^{85}$Rb, corresponding to the separable potential approach
at a magnetic field of $162.2\,$G, as obtained from the integral 
form of the Schr\"odinger equation
\be
\label{eqB16}
|\phi_{\rm b}\ra=G_0(E_{\rm b})V|\phi_{\rm b}\ra.
\ee
The radius is given on a logarithmic scale. According to Eq.~(\ref{eqB16})
the asymptotic functional behavior of both bound state wave functions at far
relative distances of the two atoms is determined solely through the free 
Green's function evaluated at the binding energy $E_{\rm b}$. As their 
molecular states are very weakly bound both wave functions extend far 
outside the range of their pair interaction.\\[0.2cm]
For $^4$He the estimate corresponding to a zero energy resonance in 
Eq.~(\ref{eqB03}) gives $|E_{\rm b}|/h=25486$ kHz while the formula
of Gribakin and Flambaum in Eq.~(\ref{eqB06})
yields $|E_{\rm b}|/h=26856$ kHz. The exact
binding energy of the TTY potential is $|E_{\rm b}|/h=27087$ kHz. The
comparison shows that $E_{\rm b}$ and, in turn, the bound state wave
function, are virtually completely determined by $a$ and $C_6$. 
Figure \ref{figHe2sep} compares the exact wave function of $^4$He$_2$ 
with the wave function obtained from the separable potential approach. 
The main small deviations occur in the region of the inner well 
of the TTY potential in Fig.~\ref{fig13}.\\[0.4cm]  

\begin{figure}[tb]
\begin{center}
\epsfig{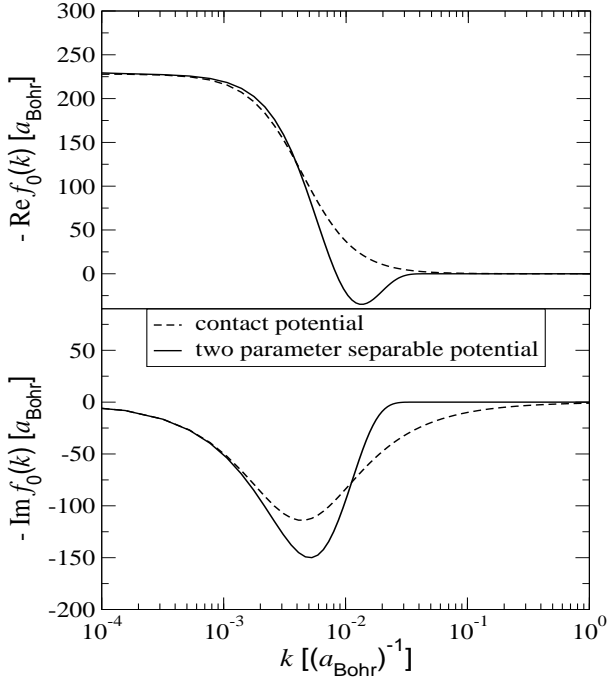}
\vspace*{5mm}
\caption{\label{fig15}The real and imaginary part of the $s$-wave scattering
amplitude $f_0(k)$ for $^{85}$Rb.
The solid lines are obtained with the two parameter separable potential, the
dashed lines in the contact potential approximation, both with a scattering
length of $a(162.2\,$G$)=228\,a_{\mathrm Bohr}$. The wave number $k$ is given
on a logarithmic scale.}
\end{center}
\end{figure}

\begin{figure}[tb]
\begin{center}
\epsfig{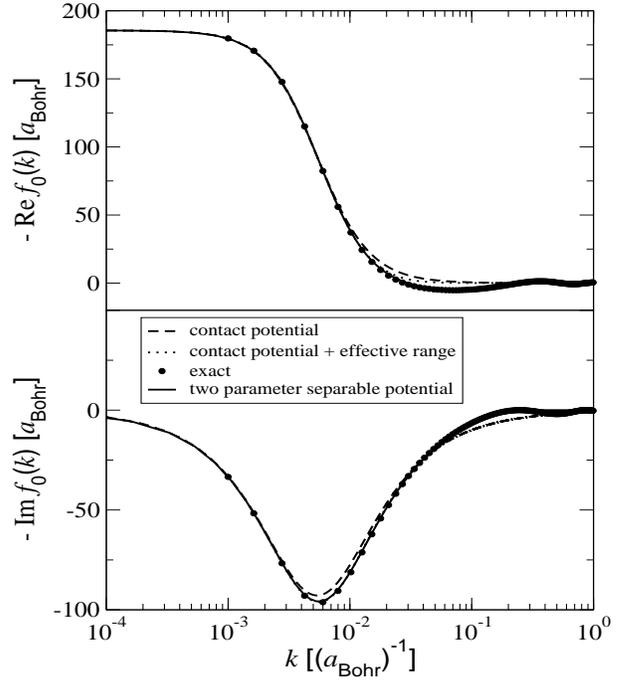}
\vspace*{5mm}
\caption{\label{fig16}The real and imaginary part of the $s$-wave scattering
amplitude $f_0(k)$ for $^{4}$He.
The solid lines are obtained with the two parameter separable potential, the
dashed lines in the contact potential approximation. The dotted lines
show the scattering amplitude of an improved contact potential approach 
\protect\cite{Braaten01}
that accounts for the effective range of the TTY potential. 
The bullets show the exact $s$ wave scattering amplitude for the TTY potential 
\protect\cite{Tang95}.}
\end{center}
\end{figure}

The $T$ matrix determines the stationary scattering wave functions that
correspond to the relative momentum $\pv$ through the Lippmann-Schwinger
equation \cite{Newton82}
\be
\label{eqB17}
|\phi_{\pv}^{(+)}\ra=|\pv\ra+G_0(p^2/m+i0)T_{\rm 2B}(p^2/m+i0)|\pv\ra,
\ee
where the energy arguments ``$p^2/m+i0$'' indicate that the real energy 
$p^2/m$ is approached from the upper half of the complex plane.
At low collision momenta $p=\hbar k$ the stationary scattering states assume 
the asymptotic form
\be
\label{eqB18}
\phi_{\pv}^{(+)}(\rv)\sim \frac{1}{\sqrt{2\pi\hbar}^3}
\left[
e^{i\pv\cdot\rv/\hbar}+f_0(k)\frac{e^{ipr/\hbar}}{r}
\right],
\ee
as soon as the relative distance $r$ exceeds by far the range of the potential.
Figure \ref{fig15} compares the $s$ wave scattering amplitude in 
Eq.~(\ref{eqB18}), for two $^{85}$Rb atoms, obtained from the separable
potential given by Eq.~(\ref{eqB13}), at $B=162.2$ G, with the amplitude of 
the contact potential in Eq.~(\ref{eqB01}).\\
The pronounced deviations at $k > 10^{-3}\ a_{\rm Bohr}^{-1}$ are related
to the large van der Waals length of the $^{85}$Rb interaction. This length
scale is not accounted for by Eq.~(\ref{eqB01}). The $s$ wave scattering 
amplitude approaches Eq.~(\ref{eqB01}) once the magnetic field is shifted
further toward the Feshbach resonance at $B=154.9$ G. The analogous comparison
for $^4$He in Fig.~\ref{fig16} may illustrate to which degree of accuracy
the scattering from the long range part of the binary interaction is 
described by the separable potential given by Eq.~(\ref{eqB13}).
In accordance with the small van der Waals length of helium of about
$10\ a_{\rm Bohr}$ the deviations between the contact potential approach
in Eq.~(\ref{eqB01}) and the exact scattering amplitude are much less 
pronounced. Even the small deviations, however, are correctly accounted for
in the separable potential approach up to wave numbers of about
$3 \times 10^{-1}\ a_{\rm Bohr}^{-1}$. The length scale related to this
upper limit of the wave numbers roughly corresponds to the radius of the
inner well of the TTY potential in Fig.~\ref{fig13}.\\[5mm]
%
%
\subsection{Dynamics}
\label{secB.2}
\noindent
In Subsection \ref{secB.1} we have analyzed the static low energy scattering
properties of two $^{85}$Rb atoms at a given magnetic field. In this
subsection we shall determine the collision dynamics that enters the 
the many body theory of a partially condensed Bose gas through coupling
functions of the form of Eq.~(\ref{eq3.9}). These coupling functions 
involve the complete unitary time evolution operator of two $^{85}$Rb atoms, 
$U_{\rm 2B}(t,\tau)$, exposed to a magnetic field pulse as shown in 
Fig.~\ref{fig1}. We shall apply the separable potential approach of 
Subsection \ref{secB.1} to determine the coupling functions 
as the effective low 
energy potential renders the time dependent Schr\"odinger equation into a 
practical form. 

We shall first determine the coupling function of the non-linear Schr\"odinger
Eq.~(\ref{eq2.20}) denoted as $h(t,\tau)$ in Eq.~(\ref{eq3.9}).
The coupling function $h(t,\tau)$
can be represented in terms of the time developed zero momentum plane wave
of the relative motion of two atoms, 
$|\zeta(t)\ra=U_{\rm 2B}(t,\tau)|0\ra$, in the form:
\be
\label{eq3.31}
h(t,\tau)=\theta(t-\tau)(2\pi\hbar)^3\la 0|V(t)|\zeta(t)\ra.
\ee
The wave function $|\zeta(t)\ra$ is determined by the integral form of the 
time dependent Schr\"odinger equation through
\be
  |\,\zeta(t)\ra= |\,0\ra + \int_\tau^t d\tau'\,
    G_0(t-\tau')V(\tau')\,|\,\zeta(\tau')\ra,
\label{eq3.32}
\ee
where $G_0(t)=\theta(t)U_0(t)/i\hbar$ is the two body Green's function of 
the relative motion of two non-interacting atoms. To obtain the coupling
function through Eq.~(\ref{eq3.31}), on the basis of the actual 
binary potential
$V(t)$, the Schr\"odinger Eq.~(\ref{eq3.32}) needs to be solved for all times 
$(t,\tau)$ between the initial and final time of the magnetic field pulse
and, moreover, at all relative distances in the argument of the wave function 
$\zeta(\rv,t)$.\\[3mm]
\begin{figure}[tb]
\begin{center}
\epsfig{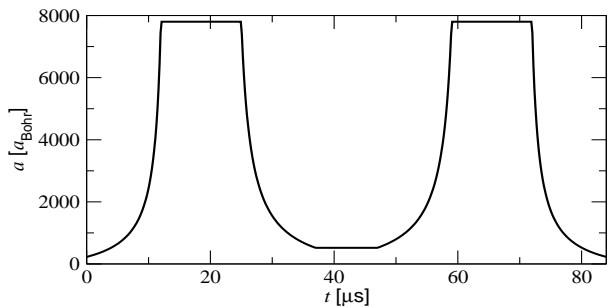}
\vspace*{5mm}
\caption{\label{fig17}The variation of the $s$ wave scattering
length $a$ corresponding to the magnetic field pulse in Fig.~\ref{fig1}.}
\end{center}
\end{figure}

The magnetic field pulse, however, releases a sufficiently small amount of 
energy to the gas that the actual potential $V(t)$ in Eq.~(\ref{eq3.32})
can be replaced by the effective low energy potential $V_{\rm sep}$ 
in Eq.~(\ref{eqB12}). Thereby, the amplitude $\xi=\xi(t)$ accounts for the
time dependence of the magnetic field through the variation of the scattering 
length $a$ illustrated in Fig.~\ref{fig17}. 
Equations (\ref{eq3.31}) and (\ref{eq3.32}) then yield the closed integral 
equation
\bea
  &&h(t,\tau)
  = (2\pi\hbar)^3|\la0\,|\,\chi\ra|^2\xi(t,\tau)
  \nonumber\\
  &&\ + \ \xi(t,\tau)\int_{\tau}^t d\tau'\,
    \la\chi\,|\,G_0(t-\tau')\,|\,\chi\ra\,h(\tau',\tau),
    \label{eq3.33}
\eea
where $\xi(t,\tau)\equiv\xi(t)\theta(t-\tau)$. The separable form of
the effective low energy potential leads to a closed dynamic equation
for $h(t,\tau)$ which avoids to explicitly take into account the spatial
dependence of $\zeta(\rv,t)$. 
The coupling function $h(t,\tau)$, as obtained from Eq.~(\ref{eq3.33}), 
is shown in Figs.~\ref{fig3} and \ref{fig5}.

\begin{figure}[tb]
\begin{center}
\epsfig{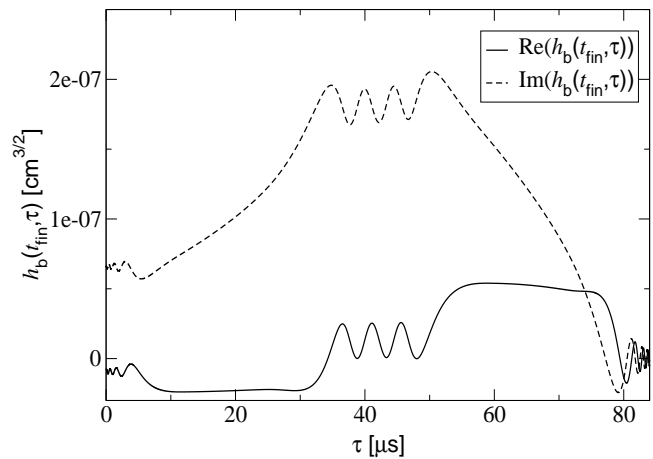}
\vspace*{5mm}
\caption{\label{fig19}The molecular coupling function $h_b(t_{\rm fin},\tau)$, 
as a function of $\tau$, corresponding to the molecular wave function of
two $^{85}$Rb atoms at the final time of the pulse as depicted in 
Fig.~\ref{fig13}. The magnetic field pulse corresponds to 
Fig.~\ref{fig1} with $t_{\rm evolve}=10\ \mu$s.}
\end{center}
\end{figure}

The coupling function associated with the molecular condensate wave function 
$\Psi_{\rm b}$ is given in Appendix \ref{appA} by Eq.~(\ref{eq4.6}) and
denoted as $h_{\rm b}(t,\tau)$.
In Section \ref{sec3} $|\Psi_{\rm b}|^2$ describes the
density of $^{85}$Rb$_2$ molecules at time $t_{\rm fin}$, 
immediately after the magnetic field pulse, in the 
bound state corresponding to the wave function in Fig.~\ref{fig13}.
The wave function $\zeta(\rv,t_{\rm fin})$ in Eq.~(\ref{eq3.32}) determines
$h_{\rm b}(t_{\rm fin},\tau)$ through:
\be
\label{eq3.34}
h_{\rm b}(t_{\rm fin},\tau)=
\theta(t_{\rm fin}-\tau)(2\pi\hbar)^{3/2}\la \phi_{\rm b}
|\zeta(t_{\rm fin})\ra.
\ee
Taking advantage of the separable form of the effective low energy potential 
the Schr\"odinger Eq.~(\ref{eq3.32}) inserted into Eq.~(\ref{eq3.34})  
determines $h_{\rm b}(t_{\rm fin},\tau)$ 
in terms of the known coupling function
$h(t,\tau)$: 
\bea
\nonumber
&&h_{\rm b}(t_{\rm fin},\tau)=\theta(t_{\rm fin}-\tau)
\bigg[(2\pi\hbar)^{3/2}\la \phi_{\rm b}|0\ra\\
&&\qquad\qquad +
\int_\tau^{t_{\rm fin}} dt
\frac{\la\phi_{\rm b}|G_0(t_{\rm fin}-t)|\chi\ra}{\la 0|\chi\ra}
h(t,\tau)\bigg].
\label{eq3.35}
\eea
The molecular coupling function $h_{\rm b}(t_{\rm fin},\tau)$, as a function
of $\tau$, is shown in Fig.~\ref{fig19} for the magnetic field pulse in 
Fig.~\ref{fig1}. The calculation of $h_{\rm b}(t_{\rm fin},\tau)$
has been performed with the $^{85}$Rb$_2$ 
wave function in Fig.~\ref{fig13} that corresponds to the shallow $s$ wave 
bound state at the magnetic field of $B=162.2$ G at the end of the pulse 
sequence.\\
%
%
A relation, similar to Eq.~(\ref{eq3.35}), with $\phi_{\rm b}$ replaced 
by the stationary 
scattering state $\phi_{\pv}^{(+)}$ has been applied to calculate
the spectral density of the pairs of burst atoms at a relative
kinetic energy $E_{\rm rel}=p^2/m$ in a homogeneous gas in 
Subsection \ref{sec4.1.1}. 
\end{appendix}

\end{multicols}
\end{document}